\documentclass[%
 reprint,
 amsmath,amssymb,
 aps,prl,
]{revtex4-2}

\usepackage{graphicx}
\usepackage{dcolumn}
\usepackage{bm}
\usepackage{hyperref}

\usepackage{color}
\usepackage{cancel}
\usepackage{physics}

\newcommand{\av}[1]{\left\langle #1 \right\rangle}

\newcommand{\FP}{\text{FP}}
\newcommand{\meas}{\text{m}}
\newcommand{\ex}{\text{ex}}
\newcommand{\hk}{\text{hk}}
\newcommand{\sts}{\text{s}}

\newcommand{\deltaFP}{\Delta_{\FP}}
\newcommand{\deltam}{\Delta_{\meas}}

\newcommand{\Px}{P(x,t)}
\newcommand{\Pxb}{P(x,t_n^-)}
\newcommand{\Pxa}{P(x,t_n^+)}

\newcommand{\Pxc}{P(x,c,t)}

\newcommand{\Pxca}{P(x,c,t_n^+)}

\newcommand{\Pcfx}{P(c|x,t)}

\newcommand{\Pxfc}{P(x|c,t)}
\newcommand{\Psxfc}{P_{\sts}(x|c)}

\newcommand{\J}{J(x,c,t)}

\newcommand{\Jtwo}{J^2(x,c,t)}

\newcommand{\calX}{\mathcal{X}}

\newcommand{\calI}{\mathcal{I}}
\newcommand{\calD}{\mathcal{D}}

\newcommand{\revision}[1]{\textcolor{black}{#1}}

\begin{document}


\title{Entropic balance with feedback control: information equalities and tight inequalities}

\author{N.~Ruiz-Pino}
\email{nruiz1@us.es}
\author{A.~Prados}%
 \email{prados@us.es}
\affiliation{%
 Física Teórica, Multidisciplinary Unit for Energy Science, Universidad de Sevilla, Apdo.~de Correos 1065, E-41080 Sevilla, Spain
}%

\date{\today}

\begin{abstract}
We consider overdamped physical systems evolving under a feedback-controlled fluctuating potential and in contact with a thermal bath at temperature $T$. A Markovian description of the dynamics, which keeps only the last value of the control action, is advantageous---both from the theoretical and the practical side---for the entropy balance. Novel second-law equalities and bounds for the extractable work are obtained, the latter being both tighter and easier to evaluate than those in the literature based on the whole chain of controller actions. The Markovian framework also allows us to prove that the bound for the extractable work that incorporates the unavailable information is saturated in a wide class of physical systems, for error-free measurements.  These results are illustrated in \revision{model systems}. For imperfect measurements, there appears an interval of measurement uncertainty, including the point at which work ceases to be extracted, where the new Markovian bound is tighter than the unavailable information bound.
\end{abstract}

\maketitle


Control of the time evolution of a system---i.e. steering it as desired---is a common objective in many contexts~\cite{bechhoefer_feedback_2005}, using either ``open-loop'' (feedforward) or ``closed-loop'' (feedback) protocols—see Refs.~\cite{guery-odelin_driving_2023,alvarado_optimal_2025,goerlich_experimental_2025} for recent reviews. Feedback control involves measuring the system state and tuning parameters accordingly to achieve a desired outcome---such as a directed flux between measurements, which can be used to perform work against an external force, as in ratchets~\cite{cao_feedback_2004,cao_thermodynamics_2009,admon_experimental_2018,saha_maximizing_2021,saha_bayesian_2022,ruiz-pino_information_2023}, or to extract work by reducing the particle’s energy at the measurement~\cite{toyabe_experimental_2010,abreu_thermodynamics_2012,ashida_general_2014,toyabe_nonequilibrium_2015,paneru_lossless_2018,paneru_optimal_2018,paneru_efficiency_2020}.

Feedback systems---from theoretical constructs like Maxwell’s demon~\cite{thomson_sorting_1879} or Szilard's engine~\cite{szilard_uber_1929} to natural systems such as molecular motors---challenge the reconciliation of the second law of thermodynamics with the role of information. The coarse-graining involved in modelling the controller---or ``demon''---makes it necessary to rewrite the second law to include an information term linking system and demon variables~\cite{schreiber_measuring_2000,sagawa_second_2008,cao_information_2009,cao_thermodynamics_2009,sagawa_generalized_2010,horowitz_nonequilibrium_2010,ponmurugan_generalized_2010,sagawa_nonequilibrium_2012,sagawa_thermodynamics_2012,abreu_thermodynamics_2012,seifert_stochastic_2012,lahiri_fluctuation_2012,sagawa_role_2013,horowitz_second-law-like_2014,ashida_general_2014,parrondo_thermodynamics_2015}.

Depending on the system and controller details, different generalisations of the second law arise.
Beyond their theoretical relevance---e.g.~in relation to the Jarzynski equality~\cite{jarzynski_nonequilibrium_1997}, they provide bounds on the extractable work. Different controller models yield different second-law variants, as each represents a distinct demon implementation with its own energy cost \cite{horowitz_second-law-like_2014}.

For a given 
protocol, the utility of these second-law variants depends on how easily the associated information function can be computed and how tight the bound on extractable work is. In some cyclic engines, like Szilard's, a simplifying feature is system and controller states \revision{becoming} independent at the end of each cycle. This reduces the thermodynamic balance to \revision{a single measurement, with} mutual information being key~\cite{sagawa_generalized_2010}. \revision{Yet}, in more complex systems like feedback flashing ratchets, system and controller states remain correlated across cycles~\cite{sagawa_nonequilibrium_2012}. One approach is to model the controller as having an infinite memory that stores all the control history,
where transfer entropy is key~\cite{cao_thermodynamics_2009, horowitz_nonequilibrium_2010, sagawa_nonequilibrium_2012,abreu_thermodynamics_2012,lahiri_fluctuation_2012,admon_experimental_2018,ruiz-pino_information_2023}.

These second-law variants have significantly advanced our understanding of measurement and feedback in physical systems, but crucial challenges remain. First, the tightness of the bounds on extractable work is not fully established~\footnote{In a specific Brownian information engine, the bound involving the unavailable information has been shown to become an equality~\cite{paneru_lossless_2018,archambault_inertial_2024}.}. Second, computing quantities like the transfer entropy is challenging, both analytically and numerically, due to correlations 
between measurements~\footnote{Thus, the transfer entropy has only been computed for very simple systems. For error-free measurements, recent work suggests that it can be estimated with moderately long chains~\cite{ruiz-pino_information_2023}.}. This Letter addresses the above limitations, both fundamental and practical, in the state of the art.

Our theoretical framework is a Markovian description of the \revision{system and controller} joint stochastic process~\cite{ruiz-pino_markovian_2024}, where only the most recent control value is kept---even \revision{for correlated measurements.}
The Markovian entropy balance 
offers both theoretical and practical advantages. It yields a new information bound for extractable work that is both tighter and easier to evaluate than the bound based on the full control history~\cite{cao_thermodynamics_2009, horowitz_nonequilibrium_2010, sagawa_nonequilibrium_2012,abreu_thermodynamics_2012,lahiri_fluctuation_2012,admon_experimental_2018,ruiz-pino_information_2023}. Also, the Markovian framework allows us to prove that the second-law inequality involving the so-called unavailable information~\cite{ashida_general_2014} becomes an equality for a broad class of systems. All results are illustrated using \revision{model systems}, including a comparison between the new Markovian bound and the unavailable information bound.

\textit{General Markovian framework--} We consider a general class of physical systems, described by overdamped Brownian motion with friction $\gamma$ in a thermal bath at temperature $T=\beta^{-1}$~\footnote{We take $k_B=1$ throughout the paper.}. The system state $x$ evolves under the action of a conservative force, deriving from a fluctuating potential $V(x,c)$, and an external force $f$; the total force is thus $F(x,c)=-\partial_{x}V(x,c)+f$. The variable $c$ describes the fluctuation of the potential, $V(x,c)\ne V(x,c')$ if $c\ne c'$, and is set by a feedback mechanism~\footnote{\revision{The variable $c$ may be multidimensional, we only assume that its set of possible values is countable.}}. In the following, we summarise the key points of the Markovian framework that are necessary for our analysis, see~\cite{ruiz-pino_markovian_2024} and Sec.~I of~\cite{SM-Markovian-entropy-balance-2025} for more details.

\begin{figure}[t]
    \centering
    \includegraphics[width=3in]{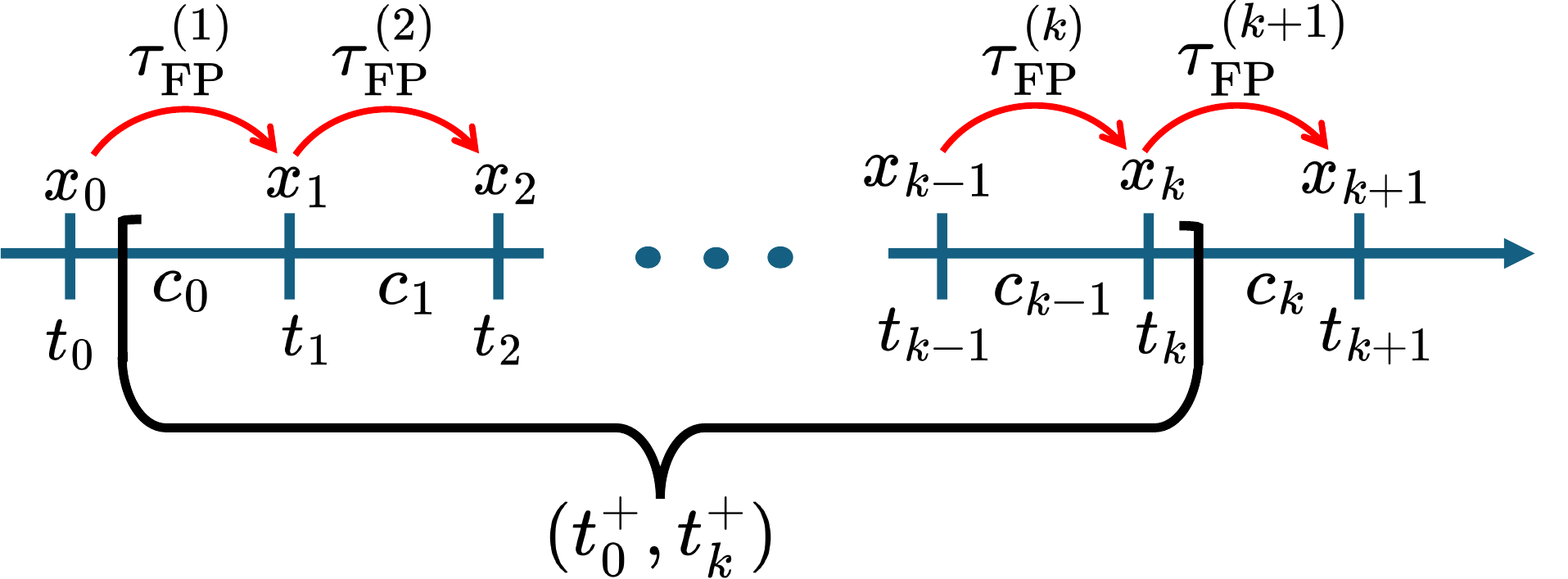}
    \caption{Sketch of the system dynamics. The feedback controller measures the system state at regular times $t_n=n\Delta t_{\meas}$, $x_n\equiv x(t_n)$, $n=0,1,\ldots$ The value of the controller variable $c_{n}$ in the time interval $\tau_{\FP}^{(n+1)}\equiv (t_{n}^+,t_{n+1}^-)$ is drawn from a probability distribution $\Theta(c_{n}|x_{n})$, conditioned to the measured $x_{n}$. The system follows overdamped Fokker-Planck (FP) evolution under a fluctuating potential $V(x,c_{n})$, different for each value  of $c_n$, in the same interval. 
    }
    \label{fig:interval}
\end{figure}
Figure~\ref{fig:interval} sketches the system evolution. Between measurements: (i) in the trajectory picture, the control variable equals the value $c_{n}$ stemming from the last measurement $x_n$, (ii) in the ensemble picture, the joint probability $\Pxc$ obeys the FP equation with potential $V(x,c)$~\cite{risken_fokker-planck_1996,gardiner_stochastic_2009}: $\partial_t \Pxc=-\partial_x \J$, with $-\gamma \J\equiv -F(x,c) \, \Pxc +  \beta^{-1}\,\partial_x \Pxc$. Since the control variable \revision{is frozen} between measurements, the conditional probability $P(x|c,t)$ also obeys the FP equation with current $J(x|c,t)$. Its normalised stationary distribution and current are $P_{\sts}(x|c)$ and $J_{\sts}(x|c)$. 
The system  variable $x$ \revision{is $d$-dimensional,} $x\in \calD\subset\mathbb{R}^d$---typically, \revision{$\calD$ is} either  $\mathbb{R}^d$, with natural boundary conditions at infinity, or a cube with periodic boundary conditions. \revision{For $d>1$, $x$, $\partial_x$, $F$, $J$ are vectors and $\partial_x J$ is the divergence of $J$.}

The  controller updates its state, in  the infinitesimal interval $\tau_{\meas}^{(n)}\equiv (t_n^-,t_n^+)$, to $c_n$ with conditional probability $\Theta(c_n|x_n)$ at $t_n^+$. Therefore, $\Pxca=\Theta(c|x) P(x,t_n)$, and $\Px$ is continuous at $t_n$, $\Pxa=\Pxb$. 
For error-free measurement, $x_n$ univocally determines $c_n$: $\calD$ splits into non-overlapping subsets $\calX_{c}$, $\cup_c \calX_c=\calD$, and \revision{the controller switches on} $V(x,c_n)$ if $x_n\in\calX_{c_n}$. Then, $\Theta(c|x)=\bm{1}_c(x)$, where $\bm{1}_c(x)$ is the indicator function of the interval $\calX_c$. \revision{The numerical section shows a specific example of $\Theta(c|x)$ for measurements with errors.}

\revision{Due to the periodic update of the feedback controller,} a time-periodic state with period $\Delta t_{\meas}$ is reached for long times, $n\ge n_R\gg 1$. Redefining $\{n\to n-n_R,t\to t-n_R \Delta t_{\meas}\}$, $n\ge 0$ and $t\ge 0$ correspond to the long-time regime. Henceforth, $P(x,c,t)$ refers to the long-time periodic solution; $P(x,t)$ and $P(c)$ to the marginal distributions---the latter is time-independent.
The chains of measured positions and corresponding control actions up to time $t_n^+$ are denoted by $\vec{x}_n=\{x_0,x_1,\ldots,x_n\}$ and $\vec{c}_n=\{c_0,c_1,\ldots\,c_n\}$, with joint probability $P(\vec{x}_n,\vec{c}_n)$. Interestingly, $\vec{x}_n$ is Markovian but $\vec{c}_n$ is not: 
\begin{equation}\label{eq:x-Markov-c-not}
   \!\!\! P(x_n|\vec{x}_{n-1})\!=\!P(x_n|x_{n-1}), \, P(c_n|\vec{c}_{n-1})\!\ne \! P(c_n|c_{n-1}). 
\end{equation} 

\textit{Thermodynamic balance--} We formulate the thermodynamic balance in the long-time periodic state. Consider the change in the average value of any quantity $A(x,c)$, $\av{A}(t)\equiv\sum_c\int dx\, A(x,c)\Pxc$, over the interval $(t_0^+,t_k^+)$ depicted in Fig.~\ref{fig:interval}, with $k$ cycles. Periodicity entails $n$-independence of: (i) the values of $\av{A}$ at the beginning and the end of $\tau_{\FP}^{(n)}$, at $t_{n-1}^+$ and $t_n^-$, so we denote them by $\av{A}^+$ and $\av{A}^-$, respectively, and (ii) its variation $\Delta\!\av{A}$ in any period $\tau_{\FP}^{(n)}\cup\tau_{\meas}^{(n)}$, which vanishes, $\Delta\!\av{A}=0$. Moreover, $\Delta\!\av{A}$ can be split into its increments $\deltaFP\av{A}$ for the FP evolution in the interval $\tau_{\FP}^{(n)}$ and $\deltam\av{A}$  at the $n$-th control update in the infinitesimal interval $\tau_{\meas}^{(n)}$:    $\Delta\!\av{A}=\deltaFP\av{A}+\deltam\av{A}=0$, $\deltaFP\av{A}\equiv\av{A}^--\av{A}^+$, $\deltam\av{A}\equiv \av{A}^+-\av{A}^-$.

The internal energy of the particle depends on both its state $x$ and the control state $c$, with average $\av{V}(t)= \sum_{c}\int dx\;  V(x,c) P(x,c,t)$. The change of $\av{V}$ during the FP evolution can be split into the heat exchanged with the thermal bath and the work done on the particle by the external force, $\deltaFP \av{V}=\av{W_{\FP}}+\av{Q}$, where
\begin{subequations}\label{eq:energy-balance-FP}
  \begin{align}
    \av{W_{\FP}}&\equiv \int_{t_{n-1}^+}^{t_{n}^-} dt \sum_c \int dx\; f J(x,c,t), \label{eq:WFP-def}\\ 
    \av{Q}&\equiv - \int_{t_{n-1}^+}^{t_{n}^-} dt \sum_c \int dx\; F(x,c) J(x,c,t).
\end{align}  
\end{subequations}
The internal energy change at the control update is the work done on the particle by the controller $\av{W_{\meas}}$, $\av{W_{\meas}}\equiv \deltam\av{V}=-\deltaFP\av{V}$. The total work is thus $\av{W}=\av{W_{\FP}}+\av{W_{\meas}}=-\av{Q}$~\footnote{We follow the standard convention for heat and work, both of them are positive (negative) when they lead to an increase (decrease) of the internal energy of the system.}.


\textit{Bounds for the work--} %
The entropy of the joint process $(x,c)$ is
$S_{xc}(t)\equiv -\sum_c \int dx\, \Pxc \ln\Pxc=S_c+S_{x|c}(t)$,
where $S_c= -\sum_c P(c) \ln P(c)$ is the time-independent control's entropy and
\begin{align}\label{eq:S-x-c} 
    S_{x|c}(t) &=-\sum_c P(c) \! \int dx \, \Pxfc \ln\Pxfc. 
\end{align}
is the time-periodic particle's entropy.

It is equivalent to analyse the dynamics of $S_{xc}(t)$ and $S_{x|c}(t)$.
In Sec.~II of~\cite{SM-Markovian-entropy-balance-2025}, we show that
\begin{align}\label{eq:deltaS-FP-and-m}
    &\deltaFP S_{x|c}=\beta \av{Q}+\Sigma_{xc}, \quad
    \deltam S_{x|c}=-\deltam\, I,
\end{align}
where the entropy production during the FP evolution is
\begin{equation}\label{eq:sigmaxc-def}
   \Sigma_{xc}=\int_{t_{n-1}^+}^{t_{n}^-} dt\, \sigma_{xc}, \quad \sigma_{xc}\equiv \gamma\beta \sum_c \int dx\, \frac{\Jtwo}{\Pxc},
\end{equation}
$\sigma_{xc}\ge 0$, and 
$\deltam I=I_{+}-I_{-}$ is the change in mutual information~\cite{cover_elements_2006,parrondo_thermodynamics_2015} between $x$ and $c$ at the measurement,
\begin{align}
  I_{\pm}=\sum_c \int dx\, P(x,c,t_n^\pm) \ln \frac{P(x|c,t_n^\pm)}{P(x,t_n)}=S_c-S_{c|x}^\pm,
\end{align}
where $S_{c|x}(t)=\sum_c\int dx\, \Pxc \ln \Pcfx$. 

\revision{The steady distribution $P_{\sts}(x|c)$ may correspond to: (i) an equilibrium state, $J_{\sts}(x|c)=0$; (ii) a non-equilibrium steady state (NESS), $J_{\sts}(x|c)\ne 0$. For periodic boundary conditions, (i) arises for $f=0$ and (ii) arises for $f\ne 0$~\footnote{\revision{For spatially periodic potentials,} the force $f$ breaks detailed balance. $J_{\sts}(x|c)\revision{\ne 0}$ is independent of $x$ \revision{for $d=1$}, \revision{whereas it} may depend on $x$ \revision{for $d>1$}.}. In case (ii),} $\Sigma_{xc}$ \revision{splits} into positive-definite housekeeping and excess contributions~\cite{oono_steady_1998,hatano_steady-state_2001,speck_integral_2005,van_den_broeck_three_2010,esposito_three_2010,seifert_stochastic_2012},
\begin{subequations}\label{eq:sigma-hk-and-ex}
  \begin{align}
    \Sigma_{xc}^{\hk}& \equiv \!\gamma\beta \!\int_{t_{n-1}^+}^{t_n^-} \!\!\!dt \sum_c\!\int\! dx\frac{J_{\sts}^2(x|c)}{P_{\sts}^2(x|c)}\Pxc, 
    \label{eq:sigma-hk}\\
    \Sigma_{xc}^{\ex}& \equiv \!\gamma\beta \!\int_{t_{n-1}^+}^{t_n^-} \!\!\! dt \sum_c\! \int\! dx \left[\frac{\J}{\Pxc}-\frac{J_{\sts}(x|c)}{P_{\sts}(x|c)}\right]^2 \!\!\Pxc 
    \label{eq:sigma-ex}.
\end{align}  
\end{subequations}

Then, $\deltam\,  I=\beta \av{Q}+\Sigma_{xc}\ge \beta \av{Q}+\Sigma_{xc}^{\hk}\ge \beta \av{Q}$
and, bringing to bear the first principle $\av{W}+\av{Q}=0$,
\begin{equation}\label{eq:W-Markov-bound}
   \beta\av{W}=\Sigma_{xc}-\deltam\,  I \ge \Sigma_{xc}^{\hk}-\deltam\,  I\ge - \deltam\,  I. 
\end{equation}
\revision{For equilibrium, $\Sigma_{xc}^{\hk}=0$: all dissipation stems from feedback and its associated information flow. For a NESS, $\Sigma_{xc}^{\hk}>0$: additional dissipation stems from the housekeeping entropy production for sustaining the NESS~\cite{abreu_thermodynamics_2012}.}

Equation~\eqref{eq:W-Markov-bound} is \revision{our} first key result. \revision{The relevant regime for information engines is  $\av{W}=\av{W}_{\meas}+\av{W}_{\FP}<0$. Two limit cases are worth considering: (i) open-loop control; measurement is not used for control, thus $c_n$ is independent of $x_n$ and $S_c=S_{c|x}^{+}$;} $\deltam\,  I=-I_- <0$, \revision{(ii)} error-free closed-loop control; \revision{$c_n$ is enslaved to $x_n$ and $S_{c|x}^{+}=0$;} $\deltam\,  I=S_{c|x}^- >0$. 
\revision{Thus, feedback-control---maybe imperfect---is essential for extracting work.}

\revision{The novel inequalities~\eqref{eq:W-Markov-bound} for the extractable work}
must be compared with existing inequalities based on
the full history of control actions~\cite{cao_thermodynamics_2009,horowitz_nonequilibrium_2010,sagawa_nonequilibrium_2012,abreu_thermodynamics_2012,admon_experimental_2018,ruiz-pino_information_2023,lahiri_fluctuation_2012},
\begin{equation}\label{eq:W-chain-bound}  
\beta\av{W}\ge \Sigma_{xc}^{\hk}-\overline{\calI_{\vec{c}}}\,\ge
-\,\overline{\calI_{\vec{c}}},
\end{equation}
which is derived within our framework in Sec.~III of~\cite{SM-Markovian-entropy-balance-2025}. Therein, $\overline{\calI_{\vec{c}}}$  is the average transfer entropy of the chain of control actions over $k$ cycles, with $k\to\infty$, 
\revision{
\begin{align}\label{eq:Ic-def}
   k\,\overline{\calI_{\vec{c}}}=\!\! \sum_{n=1}^{k} \calI_{\vec{c}}^{(n)}, \;\;
   \calI_{\vec{c}}^{(n)}= -\left[ S_{x|\vec{c}}(t_n^+)-S_{x|\vec{c}}(t_n^-)\right],
\end{align}
where $S_{x|\vec{c}}$ is defined as in Eq.~\eqref{eq:S-x-c}, with $c\to \vec{c}$. Equation~\eqref{eq:Ic-def} involves calculating the entropy of a very long non-Markovian chain $\vec{c}_n$---recall Eq.~\eqref{eq:x-Markov-c-not}~\footnote{\revision{For error-free feedback control, Eq.~\eqref{eq:Ic-def} reduces to $\overline{\calI_{\vec{c}}}=S_{\vec{c}}(t_n^+)-S_{\vec{c}}(t_n^-)=S_{c_n|\vec{c}_{n-1}}\ge 0$, see also~\cite{ruiz-pino_information_2023}.}}.} In Sec.~IV of \cite{SM-Markovian-entropy-balance-2025}, we prove that $\overline{\calI_{\vec{c}}}\geq\deltam I$. Then, 
the bounds in Eq.~\eqref{eq:W-Markov-bound} are always tighter than those in \eqref{eq:W-chain-bound}:
\begin{subequations}\label{eq:comp-ineq}
 \begin{align}
    \beta\av{W}&\geq -\deltam\,  I \geq - \overline{\calI_{\vec{c}}}, \label{eq:comp-ineq-simple}\\ 
    \beta \av{W}&\geq \Sigma_{xc}^{\hk}- \deltam\,  I\geq \Sigma_{xc}^{\hk}- \overline{\calI_{\vec{c}}} \label{eq:comp-ineq-hk},
\end{align}   
\end{subequations}
Equation~\eqref{eq:comp-ineq} is the second key result of this Letter.

\textit{Work and unavailable information--} 
Our first key result, Eq.~\eqref{eq:W-Markov-bound}, is valid in the presence of an external force $f$ and for arbitrary time $\Delta t_{\meas}$ between measurements, which may have errors. Now, we will show that the equality in Eq.~\eqref{eq:W-Markov-bound} is equivalent to
\begin{equation}\label{eq:W-Ix-Iu-equality}
\beta\av{W}\!=\!\overline{\calI_u-\calI_{\vec{x}}}, \quad f=0, \; \text{error-free}, \; \forall\Delta t_{\meas},
\end{equation}
where $\calI_{\vec{x}}^{(n)}$ is the transfer entropy of the chain of measured positions $\vec{x}_n$ and $\calI_u^{(n)}$ is the corresponding unavailable information introduced in Ref.~\cite{ashida_general_2014}. The inequality derived therein, $\beta \av{W}\ge \overline{\calI_u-\calI_{\vec{x}}}$, is thus tight for a very broad class of physical systems: both the fluctuating potential $V(x,c)$ and the time interval $\Delta t_{\meas}$ are  arbitrary. Equation~\eqref{eq:W-Ix-Iu-equality} is the third key result of our Letter.

Below, we give the main steps of the derivation of Eq.~\eqref{eq:W-Ix-Iu-equality}; the full proof can be found in Sec.~V of \cite{SM-Markovian-entropy-balance-2025}. From the definition \revision{$\calI_{\vec{x}}^{(n)}\equiv S_{\vec{x}_n}-S_{\vec{x}_{n-1}}$}, we have 
\begin{align}\label{eq:Ix-double-integral}
    \calI_{\vec{x}}^{(n)}
    &=-\int dx_{n-1} dx_n P(x_{n-1},x_n)  \ln P(x_n|x_{n-1}),
\end{align}
\revision{where we have used the Markovianity of $\vec{x}_n$---as given by Eq.~\eqref{eq:x-Markov-c-not}.} For error-free measurements, $\Theta(c_i|x_i)=\bm{1}_{c_i}(x_i)$ and $P(x_n|x_{n-1})=K(x_n,\Delta t_{\meas}|x_{n-1},c_{n-1})$ if $x_{n-1}\in\calX_{c_{n-1}}$, where $K(x,t|x',c)$ is the propagator of the FP equation for the potential $V(x,c)$. Consequently, 
\begin{align}\label{eq:Ix-sum+integral}
    \calI_{\vec{x}}^{(n)}=-\sum_{c_{n-1}}\int d & x_{n-1} dx_n  P(x_{n-1},c_{n-1},x_n) \nonumber \\
    & \times \ln K(x_n,\Delta t_{\meas}|x_{n-1},c_{n-1}) ,
\end{align}
which is independent of $n$: $\calI_{\vec{x}}^{(n)}=\overline{\calI_{\vec{x}}}$, $\forall n$.

Now we \revision{look into} $\calI_u$, for 
the reverse process~\cite{ashida_general_2014}. In each trajectory of the forward process, chains $\vec{x}_k$ and $\vec{c}_k$ \revision{are generated through measurement and feedback:} $(x_0 \to c_0) \longrightarrow(x_1 \to c_1) \longrightarrow (x_2 \to c_2) \cdots \longrightarrow (x_k\to c_k)$\revision{---recall Fig.~\ref{fig:interval}}. In each trajectory of the reverse process, time is inverted but there is no feedback: in each interval $\tau_{\FP}^{(n)}$, the system moves under the same potential of the forward process, determined by $c_{n-1}$. Then, we have a fixed reverse chain of control actions, $\vec{c}_k^R\equiv\{c_0^R=c_k \to c_1^R=c_{k-1} \to \cdots c_k^R=c_{0}\}$. Also, we define reverse chains of system states $\vec{x}_n^R\equiv\{x_0^R=x_k \to x_1^R=x_{k-1} \to \cdots x_n^R=x_{k-n}\}$, generated with $\vec{c}_k^R$, $n\le k$. We denote the transition probability of reaching position $x_n^R$ under the protocol fixed by the forward process, given a prior chain of positions $\vec{x}_{n-1}^R$, by $P^R(x_n^R|\vec{x}_{n-1}^R)$. In our notation, we write $\calI_u^{(n)}=-\av{ \ln P^R(x_n^R|\vec{x}_{n-1}^R)}$, i.e.
\begin{align}\label{eq:Iu-average}
        \calI_u^{(n)}&
        = -\!\int \! d\vec{x}_k P(\vec{x}_k)\ln{ P^R(x_n^R|\vec{x}_{n-1}^R)} \nonumber \\
        &=-\!\int \! dx_{\tilde{n}}\, dx_{\tilde{n}+1} P(x_{\tilde{n}},x_{\tilde{n}+1}) \ln P^R(x_{\tilde{n}}|x_{\tilde{n}+1}),
\end{align}
where we have defined $\tilde{n}=k-n$ and taken into account that $P^R(x_n^R|\vec{x}_{n-1}^R)$ only depends on $x_{n-1}^R$: the reverse process is also Markovian, $P^R(x_n^R|\vec{x}_{n-1}^R)=P^R(x_n^R|x_{n-1}^R)=P^R(x_{\tilde{n}}|x_{\tilde{n}+1})$. The backward evolution from $x_{\tilde{n}+1}$ to $x_{\tilde{n}}$ is done with the potential $V(x,c_{\tilde{n}})$. Then,  in close analogy with Eq.~\eqref{eq:Ix-double-integral}, 
\begin{align}\label{eq:Iu-sum+integral}
    \calI_u^{(n)}
    =&-\sum_{c_{\tilde{n}}} \int dx_{\tilde{n}}\,dx_{\tilde{n}+1} P(x_{\tilde{n}},c_{\tilde{n}},x_{\tilde{n}+1}) \nonumber \\
    &\qquad\qquad\times \ln K(x_{\tilde{n}},\Delta t_{\meas}|x_{\tilde{n}+1},c_{\tilde{n}}),
\end{align}
which is also independent of $n$: $\calI_u^{(n)}=\overline{\calI_u}$, $\forall n$.

Putting together Eqs.~\eqref{eq:Ix-sum+integral} and \eqref{eq:Iu-sum+integral}---exploiting their invariance under $n$-translation, we can write
\begin{align}
    \overline{\calI_u-\calI_{\vec{x}}}=&\beta\sum_{c_{n}}\int dx_{n} dx_{n+1}  P(x_{n},c_{n},x_{n+1}) \nonumber \\
    & \qquad\qquad\times \left[V(x_n,c_n)-V(x_{n+1},c_n)\right],
\end{align}
where we have used the detailed balance condition
\begin{equation}\label{eq:detailed-balance}
    \frac{K(x_{n+1},\Delta t_{\meas}|x_{n},c_{n})}{K(x_{n},\Delta t_{\meas}|x_{n+1},c_{n})}=\frac{P_{\sts}(x_n|c_n)}{P_{\sts}(x_{n+1}|c_n)},
\end{equation}
and the canonical shape of $P_{\sts}(x|c)$. Then,
$\overline{\calI_u-\calI_{\vec{x}}}=\beta(\av{V}^+-\av{V}^-)$,
i.e. Eq.~\eqref{eq:W-Ix-Iu-equality} holds; pairs $\{x_n,c_n\}$ and $\{x_{n+1},c_n\}$ correspond to times $t_n^+$ and $t_{n+1}^-$, respectively.

For $f\ne 0$, detailed balance is not verified in general. Then,  Eq.~\eqref{eq:W-Ix-Iu-equality} ceases to be valid 
\revision{but it can be generalised for} $\Delta t_{\meas}\to\infty$, where the \revision{FP} propagator  tends to the \revision{NESS} distribution. By incorporating the housekeeping 
entropy production \revision{$\Sigma_{xc}^{\hk}$, one gets}---see Sec.~VI of~\cite{SM-Markovian-entropy-balance-2025}:
\begin{equation}\label{eq:W-Ix-Iu-equality-with-f}
\beta\av{W}=\Sigma_{xc}^{\hk}+\overline{\calI_u-\calI_{\vec{x}}}, \;\; f\ne 0, \,   \text{error-free}, \, \Delta t_{\meas} \to\infty.
\end{equation}

\textit{Application to a model information engine--} \revision{Our framework applies to a broad class of information engines, including any colloidal particle undergoing overdamped motion (negligible inertia) in a fluctuating potential controlled by a Markovian (negligible delay) feedback device, as in Refs.~\cite{toyabe_experimental_2010,toyabe_nonequilibrium_2015,paneru_lossless_2018,paneru_optimal_2018,paneru_efficiency_2020}. In this work, we focus on engines where the feedback controller aims at reducing the particle’s energy.
The goal is to maximize work extraction at the measurements~\cite{toyabe_experimental_2010,abreu_thermodynamics_2012,ashida_general_2014,toyabe_nonequilibrium_2015,paneru_lossless_2018,paneru_optimal_2018,paneru_efficiency_2020}, even at the cost of performing work during the Fokker–Planck evolution when $f\neq 0$.} 

\begin{figure}
    \centering
    \includegraphics[width=3.25in]{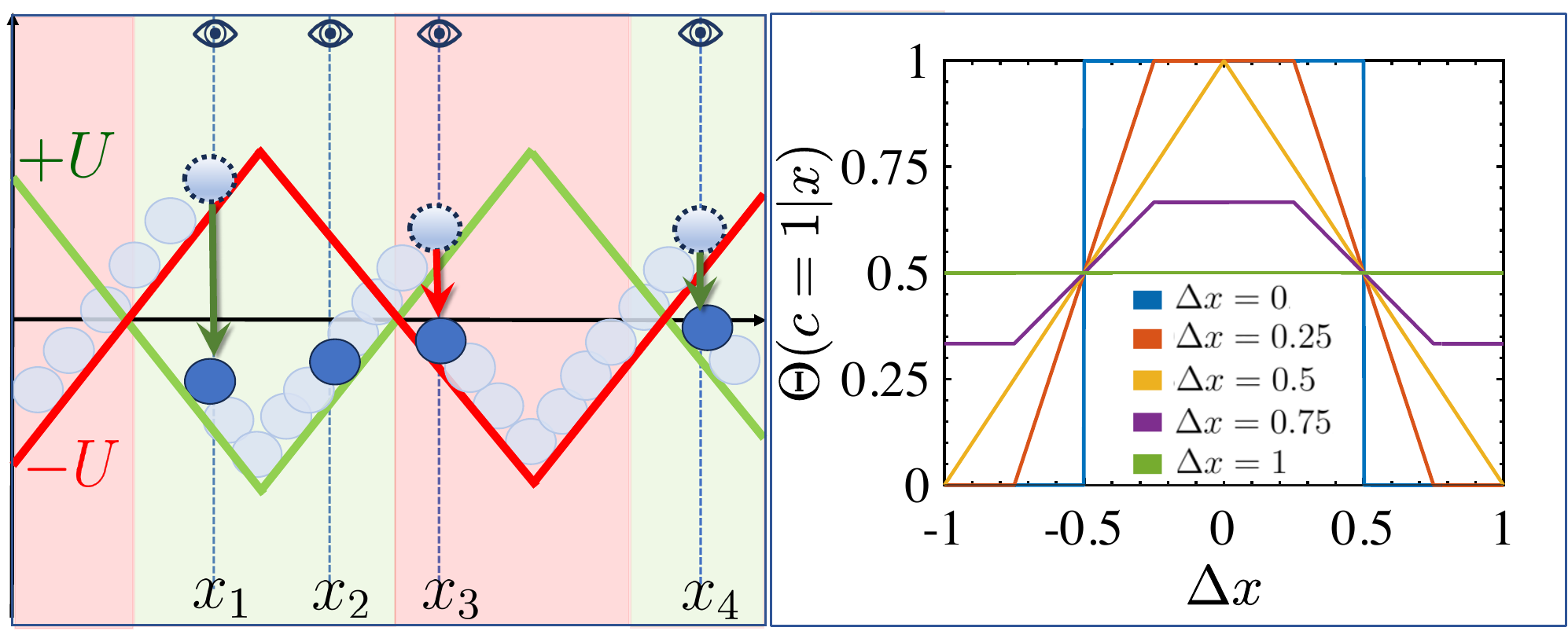}
    \caption{(Left) Sketch of an information engine. The potential may be inverted by the feedback controller after measuring the position of the particle $x_n$ at times $t_n=n\Delta t_{\meas}$. In the error-free case, the inversion is done\revision{---as shown by the arrows---}if the \revision{potential} energy \revision{decreases, so} work is always extracted. \revision{For the sake of clarity, more than one period is shown.} (Right) Introduction of error in the measurement. The particle position is measured with precision $\Delta x$. The function $\Theta(1|x)$, which gives the conditional probability of having the potential $V(x,c=1)\revision{=+U(x)}$ at $t_n^+$ is plotted for different values of $\Delta x$, from $\Delta x=0$ (error-free, 
    perfect closed-loop control) to $\Delta x=1$ (maximum error, open-loop control).}
    \label{fig:engine-types-I-and-II}
\end{figure}

\revision{Inspired by the experimental setup in Ref.~\cite{toyabe_experimental_2010},} we consider a model with a potential $U(x)$ that can be inverted by the controller, $V(x,c)=c\, U(x)$ with $c=\pm 1$. Specifically, we have considered the piecewise potential $U(x)=V_0 (|x|-L/2)/L$  in the interval $[-L,L]$ with periodic boundary conditions (left panel of Fig.~\ref{fig:engine-types-I-and-II})~\footnote{\revision{This choice preserves the physical image of a sinusoidal potential~\cite{toyabe_experimental_2010} while simplifying  analytical calculations.}}. For the error-free case,  $\Theta(c|x)=H(-c\, U(x))$, with $H(\cdot)$ being the Heaviside step function. When the particle position is measured with precision $\Delta x$, $\Theta(c|x)$ is proportional to the length of the interval $(x-\Delta x,x+\Delta x)$ with the ``correct" sign of $U(x)$ (right panel of Fig.~\ref{fig:engine-types-I-and-II}). 

\begin{figure}
    \centering
\includegraphics[width=3.35in]{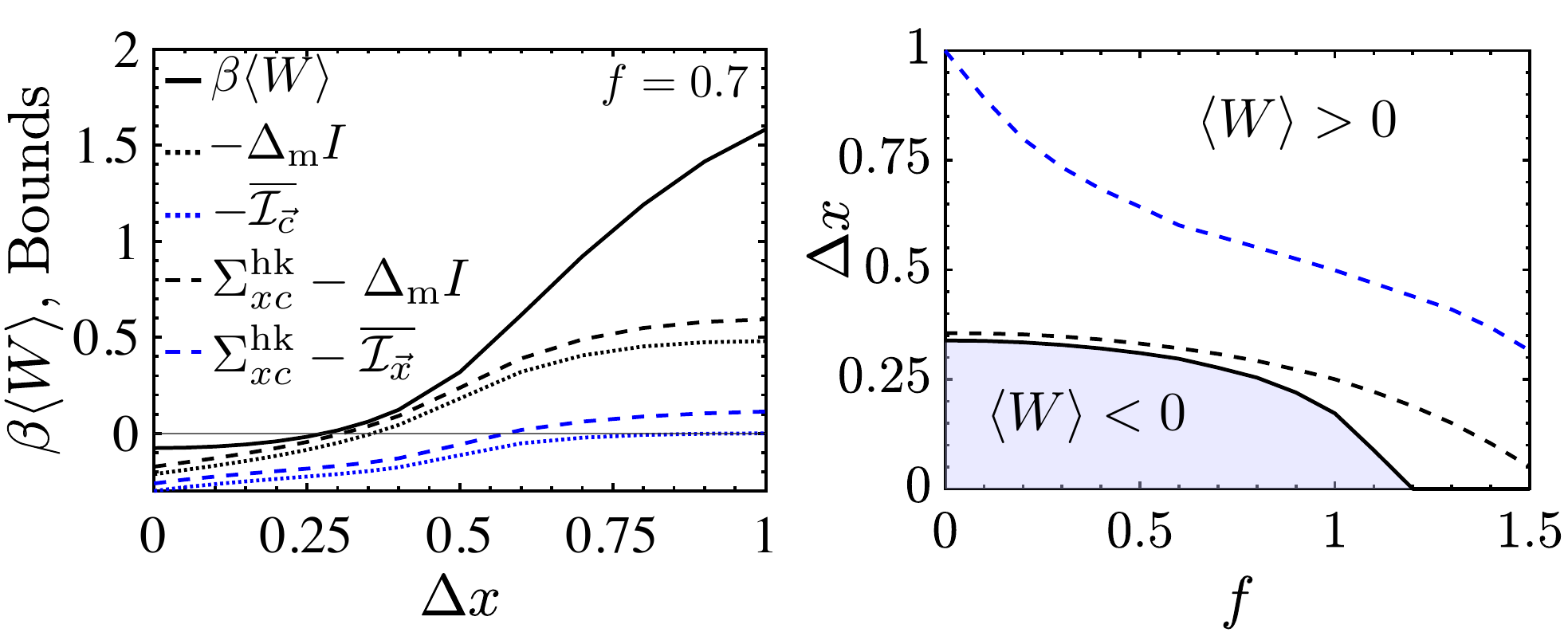}
    \caption{(Left) Average work and bounds as a function of the uncertainty in the measurement. The data shown correspond to non-equilibrium situation with $f\ne 0$, specifically $f=0.7$. The behaviour for $f=0$ is qualitatively similar, eliminating the curves with the vanishing housekeeping contribution $\Sigma_{xc}^{\hk}$.   (Right) Phase diagram of the information engine in the $(\Delta x, f)$ plane. The regions $\av{W}<0$ and $\av{W}>0$ are separated by a line $\Delta x_0(f)$ (solid), accurately predicted by the Markovian bounds (dashed black) but not by the chain ones (blue). Additional parameters are $\Delta t_{\meas}=0.5$ and $V_0=5$.
    \label{fig:type-I-num-results}
}
\end{figure}

All numerical results are in dimensionless variables: \revision{$L$ is the length unit, $T$ is the energy unit, and $\gamma L^2/T$ is the time unit}. In the left panel of Fig.~\ref{fig:type-I-num-results}, we show the average work $\av{W}$ and its bounds in Eq.~\eqref{eq:comp-ineq} as a function of $\Delta x$ for a non-equilibrium situation with $f=0.7$. As predicted by our \revision{theory}, the Markovian bounds are always tighter than their chain counterparts. It is worth noting that no net work can be extracted from the system, $\av{W}>0$, for large enough error, $\Delta x>\Delta x_0(f)$; for $\Delta x=1$, the control becomes open-loop. The value $\Delta x_0(f)$ is quite accurately predicted by the Markovian bounds, but not by the chain bounds. This is further illustrated in the right panel of Fig.~\ref{fig:type-I-num-results}, where a phase diagram of the engine in the $(\Delta x,f)$ plane is plotted.

\begin{figure}
    \centering
\includegraphics[width=3.4in]{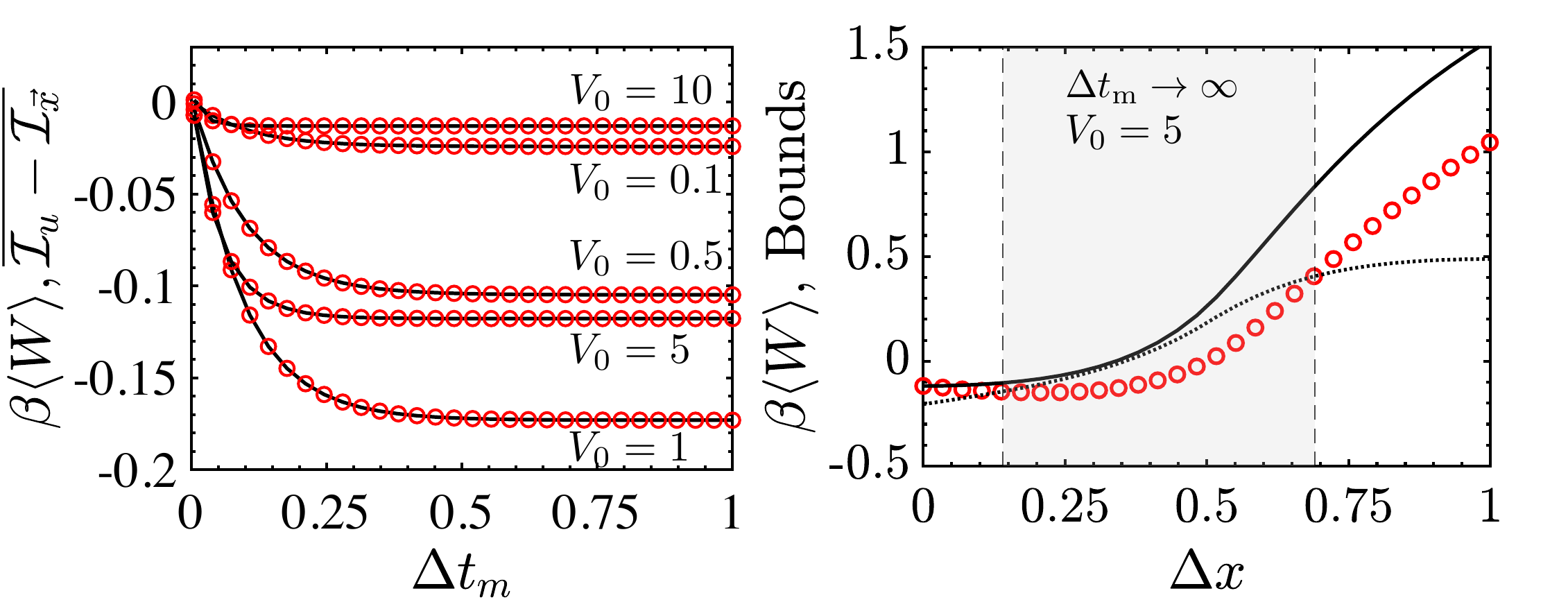}
    \caption{(Left) Numerical check of Eq.~\eqref{eq:W-Ix-Iu-equality}. Both $\beta\av{W}$ (solid line) and $\overline{\calI_u-\calI_{\vec{x}}}$ (red circles) are plotted as a function of $\Delta t_{\meas}$, for $\Delta x=0$. The agreement is excellent for all the values of $V_0$ considered, from $V_0=0.1$ ($V_0\ll T$, high temperature) to $V_0=10$ ($V_0\gg T$, low temperature). (Right) Average work and bounds as a function of $\Delta x$. Specifically, the bounds shown are the unavailable information's $\overline{\calI_u-\calI_{\vec{x}}}$ (red circles) and the Markovian's $-\deltam I$.  The region in which the Markovian bound improves the unavailable information bound is highlighted in grey. Both panels correspond to $f=0$.
    \label{fig:W-I-Iu}
}
\end{figure}

We have numerically checked equality~\eqref{eq:W-Ix-Iu-equality} in the left panel of Fig.~\ref{fig:W-I-Iu}. The extracted work $\av{W}$ is not a monotonic function of $V_0$, which can be understood intuitively: for $\Delta t_{\meas}\to\infty$, $\av{W}$ vanishes for both $V_0\to 0$ and $V_0\to\infty$; in the latter case, corresponding to low temperatures, the probability concentrates in the minimum of the potential and thus $\av{V}$ cannot be decreased at the measurement. In the right panel of Fig.~\ref{fig:W-I-Iu}, we compare the precision of the Markovian bound $-\deltam I$ and the unavailable information bound $\overline{\calI_u-\calI_{\vec{x}}}$ when there are errors in the measurement. There appears an intermediate region of $\Delta x$ in which the Markovian bound is tighter; it is inside this region that $\av{W}$ changes sign. See Sec.~VII of~\cite{SM-Markovian-entropy-balance-2025} for further numerical checks of Eqs.~\eqref{eq:W-Ix-Iu-equality} and \eqref{eq:W-Ix-Iu-equality-with-f}.

\textit{Discussion--} The improvement on the tightness of second-law inequalities stemming from the Markovian description is appealing from a conceptual point of view: the Markovian description uses the least information needed to analyse the system's dynamics by only retaining the value of the control variable determining the potential felt by the system. Also, the Markovian second-law inequalities, as given by Eq.~\eqref{eq:W-Markov-bound}, quite accurately predict the range of parameters in which model systems are able to extract work and thus are useful as information engines---\revision{for non-equilibium situations in which detailed balance is broken}, the housekeeping contribution to the entropy production \revision{should be retained}. This is a task at which previous information inequalities in the literature, stemming from the chain description, fail.

Ashida et al.\ introduced the concept of unavailable information $\overline{\calI_u}$ to improve the existing bounds for the extractable work~\cite{ashida_general_2014}. Within our Markovian modelling, we have shown analytically that the unavailable information's bound saturates for error-free measurements in a wide class of systems; Szilard's engine~\cite{ashida_general_2014} and the ``lossless'' information engine of Refs.~\cite{paneru_lossless_2018,archambault_inertial_2024}\revision{---see also End Matter---}are just particular cases, for $\Delta t_{\meas}\to\infty$, of the general result~\eqref{eq:W-Ix-Iu-equality} derived in this Letter. Furthermore, we have investigated how $\overline{\calI_u}$ behaves under imperfect feedback protocols, a setting that remained largely unexplored. Interestingly, the unavailable information bound is no longer saturated and, in fact, there exist regions of measurement error in which the Markovian bound, $-\deltam I$, improves the unavailable information's, $\overline{\calI_u-\calI_{\vec{x}}}$.

\revision{The Markovian framework accurately describes systems with feedback control in many experimental situations, when the controller's time delay is negligible against the system's typical relaxation time in contact with the thermal bath. Otherwise, the Markovian bound would be no longer applicable and one has to resort to either the transfer entropy bound or the unavailable information bound to estimate the extractable work.}

\revision{A key consideration is the choice of variables on which the information bounds depend. We have used the system state $x$ and the control variable $c$, which is updated at each measurement with probability $\Theta(c_n|x_n)$, to obtain the bound $-\deltam I(x,c)$. An alternative approach uses $x$, a measurement outcome $x'$---e.g.~$x$ plus measurement noise—and a deterministic control update $c_n = f(x'_n)$, generating the same dynamics. When only the noisy outcome $x'$ is available, the Markovian bound $-\deltam  I(x', c)$ still holds, but this bound systematically deteriorates as $x'$ becomes more coarse-grained.}

Information functions 
in the second law 
stem from the implicit coarse-graining carried out when discarding the internal state of the feedback controller in the description of the system plus controller. Introducing a microscopic modelling of the internal state of the feedback controller and its interaction with the system would make it possible to identify the microscopic origin of the information function in the mesoscopic second law. This is a relevant question that deserves to be investigated. 
On another note, the framework developed here applies to overdamped information engines. 
Extending it to the underdamped regime~\cite{dago_information_2021,dago_dynamics_2022,dago_adiabatic_2023}, which takes into account inertial effects, is another appealing perspective.

\textit{Acknowledgments--}
We acknowledge financial support from Grant PID2024-155268NB-I00 funded by MICIU/AEI/10.13039/501100011033/ FEDER, UE. We also acknowledge support from Grant ProyExcel\_00796 funded by Junta de Andalucía's PAIDI 2020 program, and the applied research and innovation Project PPIT2024-31833, cofunded by EU--Ministerio de Hacienda y Función Pública--Fondos Europeos--Junta de Andalucía--Consejería de Universidad, Investigación e Innovación. NRP acknowledges support from the FPU programme through Grant FPU21/01764.

\textit{Data Availability Statement--}
The codes employed for generating the data are openly available on the GitHub page of University of Seville's FINE research group~\cite{github-page-RyP25}.

\bibliography{Mi-biblioteca-24-jul-2025}

\begin{thebibliography}{57}%
\makeatletter
\providecommand \@ifxundefined [1]{%
 \@ifx{#1\undefined}
}%
\providecommand \@ifnum [1]{%
 \ifnum #1\expandafter \@firstoftwo
 \else \expandafter \@secondoftwo
 \fi
}%
\providecommand \@ifx [1]{%
 \ifx #1\expandafter \@firstoftwo
 \else \expandafter \@secondoftwo
 \fi
}%
\providecommand \natexlab [1]{#1}%
\providecommand \enquote  [1]{``#1''}%
\providecommand \bibnamefont  [1]{#1}%
\providecommand \bibfnamefont [1]{#1}%
\providecommand \citenamefont [1]{#1}%
\providecommand \href@noop [0]{\@secondoftwo}%
\providecommand \href [0]{\begingroup \@sanitize@url \@href}%
\providecommand \@href[1]{\@@startlink{#1}\@@href}%
\providecommand \@@href[1]{\endgroup#1\@@endlink}%
\providecommand \@sanitize@url [0]{\catcode `\\12\catcode `\$12\catcode
  `\&12\catcode `\#12\catcode `\^12\catcode `\_12\catcode `\%12\relax}%
\providecommand \@@startlink[1]{}%
\providecommand \@@endlink[0]{}%
\providecommand \url  [0]{\begingroup\@sanitize@url \@url }%
\providecommand \@url [1]{\endgroup\@href {#1}{\urlprefix }}%
\providecommand \urlprefix  [0]{URL }%
\providecommand \Eprint [0]{\href }%
\providecommand \doibase [0]{https://doi.org/}%
\providecommand \selectlanguage [0]{\@gobble}%
\providecommand \bibinfo  [0]{\@secondoftwo}%
\providecommand \bibfield  [0]{\@secondoftwo}%
\providecommand \translation [1]{[#1]}%
\providecommand \BibitemOpen [0]{}%
\providecommand \bibitemStop [0]{}%
\providecommand \bibitemNoStop [0]{.\EOS\space}%
\providecommand \EOS [0]{\spacefactor3000\relax}%
\providecommand \BibitemShut  [1]{\csname bibitem#1\endcsname}%
\let\auto@bib@innerbib\@empty
\bibitem [{\citenamefont {Bechhoefer}(2005)}]{bechhoefer_feedback_2005}%
  \BibitemOpen
  \bibfield  {author} {\bibinfo {author} {\bibfnamefont {J.}~\bibnamefont
  {Bechhoefer}},\ }\bibfield  {title} {\bibinfo {title} {Feedback for
  physicists: {A} tutorial essay on control},\ }\href
  {https://doi.org/10.1103/RevModPhys.77.783} {\bibfield  {journal} {\bibinfo
  {journal} {Reviews of Modern Physics}\ }\textbf {\bibinfo {volume} {77}},\
  \bibinfo {pages} {783} (\bibinfo {year} {2005})}\BibitemShut {NoStop}%
\bibitem [{\citenamefont {Guéry-Odelin}\ \emph {et~al.}(2023)\citenamefont
  {Guéry-Odelin}, \citenamefont {Jarzynski}, \citenamefont {Plata},
  \citenamefont {Prados},\ and\ \citenamefont
  {Trizac}}]{guery-odelin_driving_2023}%
  \BibitemOpen
  \bibfield  {author} {\bibinfo {author} {\bibfnamefont {D.}~\bibnamefont
  {Guéry-Odelin}}, \bibinfo {author} {\bibfnamefont {C.}~\bibnamefont
  {Jarzynski}}, \bibinfo {author} {\bibfnamefont {C.~A.}\ \bibnamefont
  {Plata}}, \bibinfo {author} {\bibfnamefont {A.}~\bibnamefont {Prados}},\ and\
  \bibinfo {author} {\bibfnamefont {E.}~\bibnamefont {Trizac}},\ }\bibfield
  {title} {\bibinfo {title} {Driving rapidly while remaining in control:
  classical shortcuts from {Hamiltonian} to stochastic dynamics},\ }\href
  {https://doi.org/10.1088/1361-6633/acacad} {\bibfield  {journal} {\bibinfo
  {journal} {Reports on Progress in Physics}\ }\textbf {\bibinfo {volume}
  {86}},\ \bibinfo {pages} {035902} (\bibinfo {year} {2023})}\BibitemShut
  {NoStop}%
\bibitem [{\citenamefont {Alvarado}\ \emph {et~al.}(2026)\citenamefont
  {Alvarado}, \citenamefont {Teich}, \citenamefont {Sivak},\ and\ \citenamefont
  {Bechhoefer}}]{alvarado_optimal_2025}%
  \BibitemOpen
  \bibfield  {author} {\bibinfo {author} {\bibfnamefont {J.}~\bibnamefont
  {Alvarado}}, \bibinfo {author} {\bibfnamefont {E.~G.}\ \bibnamefont {Teich}},
  \bibinfo {author} {\bibfnamefont {D.~A.}\ \bibnamefont {Sivak}},\ and\
  \bibinfo {author} {\bibfnamefont {J.}~\bibnamefont {Bechhoefer}},\ }\bibfield
   {title} {\bibinfo {title} {Optimal control in soft and active matter},\
  }\bibfield  {journal} {\bibinfo  {journal} {Annual Review of Condensed Matter
  Physics}\ }\href
  {https://doi.org/https://doi.org/10.1146/annurev-conmatphys-031324-031350}
  {https://doi.org/10.1146/annurev-conmatphys-031324-031350} (\bibinfo {year}
  {2026})\BibitemShut {NoStop}%
\bibitem [{\citenamefont {Goerlich}\ \emph {et~al.}(2025)\citenamefont
  {Goerlich}, \citenamefont {Hoek}, \citenamefont {Chor}, \citenamefont
  {Rahav},\ and\ \citenamefont {Roichman}}]{goerlich_experimental_2025}%
  \BibitemOpen
  \bibfield  {author} {\bibinfo {author} {\bibfnamefont {R.}~\bibnamefont
  {Goerlich}}, \bibinfo {author} {\bibfnamefont {L.}~\bibnamefont {Hoek}},
  \bibinfo {author} {\bibfnamefont {O.}~\bibnamefont {Chor}}, \bibinfo {author}
  {\bibfnamefont {S.}~\bibnamefont {Rahav}},\ and\ \bibinfo {author}
  {\bibfnamefont {Y.}~\bibnamefont {Roichman}},\ }\bibfield  {title} {\bibinfo
  {title} {Experimental realizations of information engines: {Beyond} proof of
  concept},\ }\href {https://doi.org/10.1209/0295-5075/adbb17} {\bibfield
  {journal} {\bibinfo  {journal} {Europhysics Letters}\ }\textbf {\bibinfo
  {volume} {149}},\ \bibinfo {pages} {61001} (\bibinfo {year}
  {2025})}\BibitemShut {NoStop}%
\bibitem [{\citenamefont {Cao}\ \emph {et~al.}(2004)\citenamefont {Cao},
  \citenamefont {Dinis},\ and\ \citenamefont {Parrondo}}]{cao_feedback_2004}%
  \BibitemOpen
  \bibfield  {author} {\bibinfo {author} {\bibfnamefont {F.~J.}\ \bibnamefont
  {Cao}}, \bibinfo {author} {\bibfnamefont {L.}~\bibnamefont {Dinis}},\ and\
  \bibinfo {author} {\bibfnamefont {J.~M.~R.}\ \bibnamefont {Parrondo}},\
  }\bibfield  {title} {\bibinfo {title} {Feedback {Control} in a {Collective}
  {Flashing} {Ratchet}},\ }\href
  {https://doi.org/10.1103/PhysRevLett.93.040603} {\bibfield  {journal}
  {\bibinfo  {journal} {Physical Review Letters}\ }\textbf {\bibinfo {volume}
  {93}},\ \bibinfo {pages} {040603} (\bibinfo {year} {2004})}\BibitemShut
  {NoStop}%
\bibitem [{\citenamefont {Cao}\ and\ \citenamefont
  {Feito}(2009)}]{cao_thermodynamics_2009}%
  \BibitemOpen
  \bibfield  {author} {\bibinfo {author} {\bibfnamefont {F.~J.}\ \bibnamefont
  {Cao}}\ and\ \bibinfo {author} {\bibfnamefont {M.}~\bibnamefont {Feito}},\
  }\bibfield  {title} {\bibinfo {title} {Thermodynamics of feedback controlled
  systems},\ }\href {https://doi.org/10.1103/PhysRevE.79.041118} {\bibfield
  {journal} {\bibinfo  {journal} {Physical Review E}\ }\textbf {\bibinfo
  {volume} {79}},\ \bibinfo {pages} {041118} (\bibinfo {year}
  {2009})}\BibitemShut {NoStop}%
\bibitem [{\citenamefont {Admon}\ \emph {et~al.}(2018)\citenamefont {Admon},
  \citenamefont {Rahav},\ and\ \citenamefont
  {Roichman}}]{admon_experimental_2018}%
  \BibitemOpen
  \bibfield  {author} {\bibinfo {author} {\bibfnamefont {T.}~\bibnamefont
  {Admon}}, \bibinfo {author} {\bibfnamefont {S.}~\bibnamefont {Rahav}},\ and\
  \bibinfo {author} {\bibfnamefont {Y.}~\bibnamefont {Roichman}},\ }\bibfield
  {title} {\bibinfo {title} {Experimental {Realization} of an {Information}
  {Machine} with {Tunable} {Temporal} {Correlations}},\ }\href
  {https://doi.org/10.1103/PhysRevLett.121.180601} {\bibfield  {journal}
  {\bibinfo  {journal} {Physical Review Letters}\ }\textbf {\bibinfo {volume}
  {121}},\ \bibinfo {pages} {180601} (\bibinfo {year} {2018})}\BibitemShut
  {NoStop}%
\bibitem [{\citenamefont {Saha}\ \emph {et~al.}(2021)\citenamefont {Saha},
  \citenamefont {Lucero}, \citenamefont {Ehrich}, \citenamefont {Sivak},\ and\
  \citenamefont {Bechhoefer}}]{saha_maximizing_2021}%
  \BibitemOpen
  \bibfield  {author} {\bibinfo {author} {\bibfnamefont {T.~K.}\ \bibnamefont
  {Saha}}, \bibinfo {author} {\bibfnamefont {J.~N.~E.}\ \bibnamefont {Lucero}},
  \bibinfo {author} {\bibfnamefont {J.}~\bibnamefont {Ehrich}}, \bibinfo
  {author} {\bibfnamefont {D.~A.}\ \bibnamefont {Sivak}},\ and\ \bibinfo
  {author} {\bibfnamefont {J.}~\bibnamefont {Bechhoefer}},\ }\bibfield  {title}
  {\bibinfo {title} {Maximizing power and velocity of an information engine},\
  }\href {https://doi.org/10.1073/pnas.2023356118} {\bibfield  {journal}
  {\bibinfo  {journal} {Proceedings of the National Academy of Sciences}\
  }\textbf {\bibinfo {volume} {118}},\ \bibinfo {pages} {e2023356118} (\bibinfo
  {year} {2021})}\BibitemShut {NoStop}%
\bibitem [{\citenamefont {Saha}\ \emph {et~al.}(2022)\citenamefont {Saha},
  \citenamefont {Lucero}, \citenamefont {Ehrich}, \citenamefont {Sivak},\ and\
  \citenamefont {Bechhoefer}}]{saha_bayesian_2022}%
  \BibitemOpen
  \bibfield  {author} {\bibinfo {author} {\bibfnamefont {T.~K.}\ \bibnamefont
  {Saha}}, \bibinfo {author} {\bibfnamefont {J.~N.}\ \bibnamefont {Lucero}},
  \bibinfo {author} {\bibfnamefont {J.}~\bibnamefont {Ehrich}}, \bibinfo
  {author} {\bibfnamefont {D.~A.}\ \bibnamefont {Sivak}},\ and\ \bibinfo
  {author} {\bibfnamefont {J.}~\bibnamefont {Bechhoefer}},\ }\bibfield  {title}
  {\bibinfo {title} {Bayesian {Information} {Engine} that {Optimally}
  {Exploits} {Noisy} {Measurements}},\ }\href
  {https://doi.org/10.1103/PhysRevLett.129.130601} {\bibfield  {journal}
  {\bibinfo  {journal} {Physical Review Letters}\ }\textbf {\bibinfo {volume}
  {129}},\ \bibinfo {pages} {130601} (\bibinfo {year} {2022})}\BibitemShut
  {NoStop}%
\bibitem [{\citenamefont {Ruiz-Pino}\ \emph {et~al.}(2023)\citenamefont
  {Ruiz-Pino}, \citenamefont {Villarrubia-Moreno}, \citenamefont {Prados},\
  and\ \citenamefont {Cao-García}}]{ruiz-pino_information_2023}%
  \BibitemOpen
  \bibfield  {author} {\bibinfo {author} {\bibfnamefont {N.}~\bibnamefont
  {Ruiz-Pino}}, \bibinfo {author} {\bibfnamefont {D.}~\bibnamefont
  {Villarrubia-Moreno}}, \bibinfo {author} {\bibfnamefont {A.}~\bibnamefont
  {Prados}},\ and\ \bibinfo {author} {\bibfnamefont {F.~J.}\ \bibnamefont
  {Cao-García}},\ }\bibfield  {title} {\bibinfo {title} {Information in
  feedback ratchets},\ }\href {https://doi.org/10.1103/PhysRevE.108.034112}
  {\bibfield  {journal} {\bibinfo  {journal} {Physical Review E}\ }\textbf
  {\bibinfo {volume} {108}},\ \bibinfo {pages} {034112} (\bibinfo {year}
  {2023})}\BibitemShut {NoStop}%
\bibitem [{\citenamefont {Toyabe}\ \emph {et~al.}(2010)\citenamefont {Toyabe},
  \citenamefont {Sagawa}, \citenamefont {Ueda}, \citenamefont {Muneyuki},\ and\
  \citenamefont {Sano}}]{toyabe_experimental_2010}%
  \BibitemOpen
  \bibfield  {author} {\bibinfo {author} {\bibfnamefont {S.}~\bibnamefont
  {Toyabe}}, \bibinfo {author} {\bibfnamefont {T.}~\bibnamefont {Sagawa}},
  \bibinfo {author} {\bibfnamefont {M.}~\bibnamefont {Ueda}}, \bibinfo {author}
  {\bibfnamefont {E.}~\bibnamefont {Muneyuki}},\ and\ \bibinfo {author}
  {\bibfnamefont {M.}~\bibnamefont {Sano}},\ }\bibfield  {title} {\bibinfo
  {title} {Experimental demonstration of information-to-energy conversion and
  validation of the generalized {Jarzynski} equality},\ }\href
  {https://doi.org/10.1038/nphys1821} {\bibfield  {journal} {\bibinfo
  {journal} {Nature Physics}\ }\textbf {\bibinfo {volume} {6}},\ \bibinfo
  {pages} {988} (\bibinfo {year} {2010})}\BibitemShut {NoStop}%
\bibitem [{\citenamefont {Abreu}\ and\ \citenamefont
  {Seifert}(2012)}]{abreu_thermodynamics_2012}%
  \BibitemOpen
  \bibfield  {author} {\bibinfo {author} {\bibfnamefont {D.}~\bibnamefont
  {Abreu}}\ and\ \bibinfo {author} {\bibfnamefont {U.}~\bibnamefont
  {Seifert}},\ }\bibfield  {title} {\bibinfo {title} {Thermodynamics of
  {Genuine} {Nonequilibrium} {States} under {Feedback} {Control}},\ }\href
  {https://doi.org/10.1103/PhysRevLett.108.030601} {\bibfield  {journal}
  {\bibinfo  {journal} {Physical Review Letters}\ }\textbf {\bibinfo {volume}
  {108}},\ \bibinfo {pages} {030601} (\bibinfo {year} {2012})}\BibitemShut
  {NoStop}%
\bibitem [{\citenamefont {Ashida}\ \emph {et~al.}(2014)\citenamefont {Ashida},
  \citenamefont {Funo}, \citenamefont {Murashita},\ and\ \citenamefont
  {Ueda}}]{ashida_general_2014}%
  \BibitemOpen
  \bibfield  {author} {\bibinfo {author} {\bibfnamefont {Y.}~\bibnamefont
  {Ashida}}, \bibinfo {author} {\bibfnamefont {K.}~\bibnamefont {Funo}},
  \bibinfo {author} {\bibfnamefont {Y.}~\bibnamefont {Murashita}},\ and\
  \bibinfo {author} {\bibfnamefont {M.}~\bibnamefont {Ueda}},\ }\bibfield
  {title} {\bibinfo {title} {General achievable bound of extractable work under
  feedback control},\ }\href {https://doi.org/10.1103/PhysRevE.90.052125}
  {\bibfield  {journal} {\bibinfo  {journal} {Physical Review E}\ }\textbf
  {\bibinfo {volume} {90}},\ \bibinfo {pages} {052125} (\bibinfo {year}
  {2014})}\BibitemShut {NoStop}%
\bibitem [{\citenamefont {Toyabe}\ and\ \citenamefont
  {Sano}(2015)}]{toyabe_nonequilibrium_2015}%
  \BibitemOpen
  \bibfield  {author} {\bibinfo {author} {\bibfnamefont {S.}~\bibnamefont
  {Toyabe}}\ and\ \bibinfo {author} {\bibfnamefont {M.}~\bibnamefont {Sano}},\
  }\bibfield  {title} {\bibinfo {title} {Nonequilibrium {Fluctuations} in
  {Biological} {Strands}, {Machines}, and {Cells}},\ }\href@noop {} {\bibfield
  {journal} {\bibinfo  {journal} {Journal of the Physical Society of Japan}\
  }\textbf {\bibinfo {volume} {84}},\ \bibinfo {pages} {102001} (\bibinfo
  {year} {2015})}\BibitemShut {NoStop}%
\bibitem [{\citenamefont {Paneru}\ \emph
  {et~al.}(2018{\natexlab{a}})\citenamefont {Paneru}, \citenamefont {Lee},
  \citenamefont {Tlusty},\ and\ \citenamefont {Pak}}]{paneru_lossless_2018}%
  \BibitemOpen
  \bibfield  {author} {\bibinfo {author} {\bibfnamefont {G.}~\bibnamefont
  {Paneru}}, \bibinfo {author} {\bibfnamefont {D.~Y.}\ \bibnamefont {Lee}},
  \bibinfo {author} {\bibfnamefont {T.}~\bibnamefont {Tlusty}},\ and\ \bibinfo
  {author} {\bibfnamefont {H.~K.}\ \bibnamefont {Pak}},\ }\bibfield  {title}
  {\bibinfo {title} {Lossless {Brownian} {Information} {Engine}},\ }\href
  {https://doi.org/10.1103/PhysRevLett.120.020601} {\bibfield  {journal}
  {\bibinfo  {journal} {Physical Review Letters}\ }\textbf {\bibinfo {volume}
  {120}},\ \bibinfo {pages} {020601} (\bibinfo {year}
  {2018}{\natexlab{a}})}\BibitemShut {NoStop}%
\bibitem [{\citenamefont {Paneru}\ \emph
  {et~al.}(2018{\natexlab{b}})\citenamefont {Paneru}, \citenamefont {Lee},
  \citenamefont {Park}, \citenamefont {Park}, \citenamefont {Noh},\ and\
  \citenamefont {Pak}}]{paneru_optimal_2018}%
  \BibitemOpen
  \bibfield  {author} {\bibinfo {author} {\bibfnamefont {G.}~\bibnamefont
  {Paneru}}, \bibinfo {author} {\bibfnamefont {D.~Y.}\ \bibnamefont {Lee}},
  \bibinfo {author} {\bibfnamefont {J.-M.}\ \bibnamefont {Park}}, \bibinfo
  {author} {\bibfnamefont {J.~T.}\ \bibnamefont {Park}}, \bibinfo {author}
  {\bibfnamefont {J.~D.}\ \bibnamefont {Noh}},\ and\ \bibinfo {author}
  {\bibfnamefont {H.~K.}\ \bibnamefont {Pak}},\ }\bibfield  {title} {\bibinfo
  {title} {Optimal tuning of a {Brownian} information engine operating in a
  nonequilibrium steady state},\ }\href
  {https://doi.org/10.1103/PhysRevE.98.052119} {\bibfield  {journal} {\bibinfo
  {journal} {Physical Review E}\ }\textbf {\bibinfo {volume} {98}},\ \bibinfo
  {pages} {052119} (\bibinfo {year} {2018}{\natexlab{b}})}\BibitemShut
  {NoStop}%
\bibitem [{\citenamefont {Paneru}\ \emph {et~al.}(2020)\citenamefont {Paneru},
  \citenamefont {Dutta}, \citenamefont {Sagawa}, \citenamefont {Tlusty},\ and\
  \citenamefont {Pak}}]{paneru_efficiency_2020}%
  \BibitemOpen
  \bibfield  {author} {\bibinfo {author} {\bibfnamefont {G.}~\bibnamefont
  {Paneru}}, \bibinfo {author} {\bibfnamefont {S.}~\bibnamefont {Dutta}},
  \bibinfo {author} {\bibfnamefont {T.}~\bibnamefont {Sagawa}}, \bibinfo
  {author} {\bibfnamefont {T.}~\bibnamefont {Tlusty}},\ and\ \bibinfo {author}
  {\bibfnamefont {H.~K.}\ \bibnamefont {Pak}},\ }\bibfield  {title} {\bibinfo
  {title} {Efficiency fluctuations and noise induced refrigerator-to-heater
  transition in information engines},\ }\href
  {https://doi.org/10.1038/s41467-020-14823-x} {\bibfield  {journal} {\bibinfo
  {journal} {Nature Communications}\ }\textbf {\bibinfo {volume} {11}},\
  \bibinfo {pages} {1012} (\bibinfo {year} {2020})}\BibitemShut {NoStop}%
\bibitem [{\citenamefont {Thomson}(1879)}]{thomson_sorting_1879}%
  \BibitemOpen
  \bibfield  {author} {\bibinfo {author} {\bibfnamefont {S.~W.}\ \bibnamefont
  {Thomson}},\ }\bibfield  {title} {\bibinfo {title} {The {Sorting} {Demon} of
  {Maxwell}},\ }\href {https://doi.org/10.1038/020126a0} {\bibfield  {journal}
  {\bibinfo  {journal} {Nature}\ }\textbf {\bibinfo {volume} {20}},\ \bibinfo
  {pages} {126} (\bibinfo {year} {1879})}\BibitemShut {NoStop}%
\bibitem [{\citenamefont {Szilard}(1929)}]{szilard_uber_1929}%
  \BibitemOpen
  \bibfield  {author} {\bibinfo {author} {\bibfnamefont {L.}~\bibnamefont
  {Szilard}},\ }\bibfield  {title} {\bibinfo {title} {über die
  {Entropieverminderung} in einem thermodynamischen {System} bei {Eingriffen}
  intelligenter {Wesen}},\ }\href {https://doi.org/10.1007/BF01341281}
  {\bibfield  {journal} {\bibinfo  {journal} {Zeitschrift für Physik}\
  }\textbf {\bibinfo {volume} {53}},\ \bibinfo {pages} {840} (\bibinfo {year}
  {1929})}\BibitemShut {NoStop}%
\bibitem [{\citenamefont {Schreiber}(2000)}]{schreiber_measuring_2000}%
  \BibitemOpen
  \bibfield  {author} {\bibinfo {author} {\bibfnamefont {T.}~\bibnamefont
  {Schreiber}},\ }\bibfield  {title} {\bibinfo {title} {Measuring {Information}
  {Transfer}},\ }\href {https://doi.org/10.1103/PhysRevLett.85.461} {\bibfield
  {journal} {\bibinfo  {journal} {Physical Review Letters}\ }\textbf {\bibinfo
  {volume} {85}},\ \bibinfo {pages} {461} (\bibinfo {year} {2000})}\BibitemShut
  {NoStop}%
\bibitem [{\citenamefont {Sagawa}\ and\ \citenamefont
  {Ueda}(2008)}]{sagawa_second_2008}%
  \BibitemOpen
  \bibfield  {author} {\bibinfo {author} {\bibfnamefont {T.}~\bibnamefont
  {Sagawa}}\ and\ \bibinfo {author} {\bibfnamefont {M.}~\bibnamefont {Ueda}},\
  }\bibfield  {title} {\bibinfo {title} {Second {Law} of {Thermodynamics} with
  {Discrete} {Quantum} {Feedback} {Control}},\ }\href
  {https://doi.org/10.1103/PhysRevLett.100.080403} {\bibfield  {journal}
  {\bibinfo  {journal} {Physical Review Letters}\ }\textbf {\bibinfo {volume}
  {100}},\ \bibinfo {pages} {080403} (\bibinfo {year} {2008})}\BibitemShut
  {NoStop}%
\bibitem [{\citenamefont {Cao}\ \emph {et~al.}(2009)\citenamefont {Cao},
  \citenamefont {Feito},\ and\ \citenamefont
  {Touchette}}]{cao_information_2009}%
  \BibitemOpen
  \bibfield  {author} {\bibinfo {author} {\bibfnamefont {F.}~\bibnamefont
  {Cao}}, \bibinfo {author} {\bibfnamefont {M.}~\bibnamefont {Feito}},\ and\
  \bibinfo {author} {\bibfnamefont {H.}~\bibnamefont {Touchette}},\ }\bibfield
  {title} {\bibinfo {title} {Information and flux in a feedback controlled
  {Brownian} ratchet},\ }\href {https://doi.org/10.1016/j.physa.2008.10.006}
  {\bibfield  {journal} {\bibinfo  {journal} {Physica A: Statistical Mechanics
  and its Applications}\ }\textbf {\bibinfo {volume} {388}},\ \bibinfo {pages}
  {113} (\bibinfo {year} {2009})}\BibitemShut {NoStop}%
\bibitem [{\citenamefont {Sagawa}\ and\ \citenamefont
  {Ueda}(2010)}]{sagawa_generalized_2010}%
  \BibitemOpen
  \bibfield  {author} {\bibinfo {author} {\bibfnamefont {T.}~\bibnamefont
  {Sagawa}}\ and\ \bibinfo {author} {\bibfnamefont {M.}~\bibnamefont {Ueda}},\
  }\bibfield  {title} {\bibinfo {title} {Generalized {Jarzynski} {Equality}
  under {Nonequilibrium} {Feedback} {Control}},\ }\href
  {https://doi.org/10.1103/PhysRevLett.104.090602} {\bibfield  {journal}
  {\bibinfo  {journal} {Physical Review Letters}\ }\textbf {\bibinfo {volume}
  {104}},\ \bibinfo {pages} {090602} (\bibinfo {year} {2010})}\BibitemShut
  {NoStop}%
\bibitem [{\citenamefont {Horowitz}\ and\ \citenamefont
  {Vaikuntanathan}(2010)}]{horowitz_nonequilibrium_2010}%
  \BibitemOpen
  \bibfield  {author} {\bibinfo {author} {\bibfnamefont {J.~M.}\ \bibnamefont
  {Horowitz}}\ and\ \bibinfo {author} {\bibfnamefont {S.}~\bibnamefont
  {Vaikuntanathan}},\ }\bibfield  {title} {\bibinfo {title} {Nonequilibrium
  detailed fluctuation theorem for repeated discrete feedback},\ }\href
  {https://doi.org/10.1103/PhysRevE.82.061120} {\bibfield  {journal} {\bibinfo
  {journal} {Physical Review E}\ }\textbf {\bibinfo {volume} {82}},\ \bibinfo
  {pages} {061120} (\bibinfo {year} {2010})}\BibitemShut {NoStop}%
\bibitem [{\citenamefont {Ponmurugan}(2010)}]{ponmurugan_generalized_2010}%
  \BibitemOpen
  \bibfield  {author} {\bibinfo {author} {\bibfnamefont {M.}~\bibnamefont
  {Ponmurugan}},\ }\bibfield  {title} {\bibinfo {title} {Generalized detailed
  fluctuation theorem under nonequilibrium feedback control},\ }\href
  {https://doi.org/10.1103/PhysRevE.82.031129} {\bibfield  {journal} {\bibinfo
  {journal} {Physical Review E}\ }\textbf {\bibinfo {volume} {82}},\ \bibinfo
  {pages} {031129} (\bibinfo {year} {2010})}\BibitemShut {NoStop}%
\bibitem [{\citenamefont {Sagawa}\ and\ \citenamefont
  {Ueda}(2012)}]{sagawa_nonequilibrium_2012}%
  \BibitemOpen
  \bibfield  {author} {\bibinfo {author} {\bibfnamefont {T.}~\bibnamefont
  {Sagawa}}\ and\ \bibinfo {author} {\bibfnamefont {M.}~\bibnamefont {Ueda}},\
  }\bibfield  {title} {\bibinfo {title} {Nonequilibrium thermodynamics of
  feedback control},\ }\href {https://doi.org/10.1103/PhysRevE.85.021104}
  {\bibfield  {journal} {\bibinfo  {journal} {Physical Review E}\ }\textbf
  {\bibinfo {volume} {85}},\ \bibinfo {pages} {021104} (\bibinfo {year}
  {2012})}\BibitemShut {NoStop}%
\bibitem [{\citenamefont {Sagawa}(2012)}]{sagawa_thermodynamics_2012}%
  \BibitemOpen
  \bibfield  {author} {\bibinfo {author} {\bibfnamefont {T.}~\bibnamefont
  {Sagawa}},\ }\bibfield  {title} {\bibinfo {title} {Thermodynamics of
  {Information} {Processing} in {Small} {Systems}},\ }\href
  {https://doi.org/10.1143/PTP.127.1} {\bibfield  {journal} {\bibinfo
  {journal} {Progress of Theoretical Physics}\ }\textbf {\bibinfo {volume}
  {127}},\ \bibinfo {pages} {1} (\bibinfo {year} {2012})}\BibitemShut {NoStop}%
\bibitem [{\citenamefont {Seifert}(2012)}]{seifert_stochastic_2012}%
  \BibitemOpen
  \bibfield  {author} {\bibinfo {author} {\bibfnamefont {U.}~\bibnamefont
  {Seifert}},\ }\bibfield  {title} {\bibinfo {title} {Stochastic
  thermodynamics, fluctuation theorems and molecular machines},\ }\href
  {https://doi.org/10.1088/0034-4885/75/12/126001} {\bibfield  {journal}
  {\bibinfo  {journal} {Reports on Progress in Physics}\ }\textbf {\bibinfo
  {volume} {75}},\ \bibinfo {pages} {126001} (\bibinfo {year}
  {2012})}\BibitemShut {NoStop}%
\bibitem [{\citenamefont {Lahiri}\ \emph {et~al.}(2012)\citenamefont {Lahiri},
  \citenamefont {Rana},\ and\ \citenamefont
  {Jayannavar}}]{lahiri_fluctuation_2012}%
  \BibitemOpen
  \bibfield  {author} {\bibinfo {author} {\bibfnamefont {S.}~\bibnamefont
  {Lahiri}}, \bibinfo {author} {\bibfnamefont {S.}~\bibnamefont {Rana}},\ and\
  \bibinfo {author} {\bibfnamefont {A.~M.}\ \bibnamefont {Jayannavar}},\
  }\bibfield  {title} {\bibinfo {title} {Fluctuation theorems in the presence
  of information gain and feedback},\ }\href
  {https://doi.org/10.1088/1751-8113/45/6/065002} {\bibfield  {journal}
  {\bibinfo  {journal} {Journal of Physics A: Mathematical and Theoretical}\
  }\textbf {\bibinfo {volume} {45}},\ \bibinfo {pages} {065002} (\bibinfo
  {year} {2012})}\BibitemShut {NoStop}%
\bibitem [{\citenamefont {Sagawa}\ and\ \citenamefont
  {Ueda}(2013)}]{sagawa_role_2013}%
  \BibitemOpen
  \bibfield  {author} {\bibinfo {author} {\bibfnamefont {T.}~\bibnamefont
  {Sagawa}}\ and\ \bibinfo {author} {\bibfnamefont {M.}~\bibnamefont {Ueda}},\
  }\bibfield  {title} {\bibinfo {title} {Role of mutual information in entropy
  production under information exchanges},\ }\href
  {https://doi.org/10.1088/1367-2630/15/12/125012} {\bibfield  {journal}
  {\bibinfo  {journal} {New Journal of Physics}\ }\textbf {\bibinfo {volume}
  {15}},\ \bibinfo {pages} {125012} (\bibinfo {year} {2013})}\BibitemShut
  {NoStop}%
\bibitem [{\citenamefont {Horowitz}\ and\ \citenamefont
  {Sandberg}(2014)}]{horowitz_second-law-like_2014}%
  \BibitemOpen
  \bibfield  {author} {\bibinfo {author} {\bibfnamefont {J.~M.}\ \bibnamefont
  {Horowitz}}\ and\ \bibinfo {author} {\bibfnamefont {H.}~\bibnamefont
  {Sandberg}},\ }\bibfield  {title} {\bibinfo {title} {Second-law-like
  inequalities with information and their interpretations},\ }\href
  {https://doi.org/10.1088/1367-2630/16/12/125007} {\bibfield  {journal}
  {\bibinfo  {journal} {New Journal of Physics}\ }\textbf {\bibinfo {volume}
  {16}},\ \bibinfo {pages} {125007} (\bibinfo {year} {2014})}\BibitemShut
  {NoStop}%
\bibitem [{\citenamefont {Parrondo}\ \emph {et~al.}(2015)\citenamefont
  {Parrondo}, \citenamefont {Horowitz},\ and\ \citenamefont
  {Sagawa}}]{parrondo_thermodynamics_2015}%
  \BibitemOpen
  \bibfield  {author} {\bibinfo {author} {\bibfnamefont {J.~M.~R.}\
  \bibnamefont {Parrondo}}, \bibinfo {author} {\bibfnamefont {J.~M.}\
  \bibnamefont {Horowitz}},\ and\ \bibinfo {author} {\bibfnamefont
  {T.}~\bibnamefont {Sagawa}},\ }\bibfield  {title} {\bibinfo {title}
  {Thermodynamics of information},\ }\href {https://doi.org/10.1038/nphys3230}
  {\bibfield  {journal} {\bibinfo  {journal} {Nature Physics}\ }\textbf
  {\bibinfo {volume} {11}},\ \bibinfo {pages} {131} (\bibinfo {year}
  {2015})}\BibitemShut {NoStop}%
\bibitem [{\citenamefont {Jarzynski}(1997)}]{jarzynski_nonequilibrium_1997}%
  \BibitemOpen
  \bibfield  {author} {\bibinfo {author} {\bibfnamefont {C.}~\bibnamefont
  {Jarzynski}},\ }\bibfield  {title} {\bibinfo {title} {Nonequilibrium
  {Equality} for {Free} {Energy} {Differences}},\ }\href
  {https://doi.org/10.1103/PhysRevLett.78.2690} {\bibfield  {journal} {\bibinfo
   {journal} {Physical Review Letters}\ }\textbf {\bibinfo {volume} {78}},\
  \bibinfo {pages} {2690} (\bibinfo {year} {1997})}\BibitemShut {NoStop}%
\bibitem [{Note1()}]{Note1}%
  \BibitemOpen
  \bibinfo {note} {In a specific Brownian information engine, the bound
  involving unavailable information has been shown to become an equality~\cite
  {paneru_lossless_2018,archambault_inertial_2024}.}\BibitemShut {Stop}%
\bibitem [{Note2()}]{Note2}%
  \BibitemOpen
  \bibinfo {note} {Thus, the transfer entropy has only been computed for very
  simple systems. For error-free measurements, recent work suggests that it can
  be estimated with moderately long chains~\cite
  {ruiz-pino_information_2023}.}\BibitemShut {Stop}%
\bibitem [{\citenamefont {Ruiz-Pino}\ and\ \citenamefont
  {Prados}(2024)}]{ruiz-pino_markovian_2024}%
  \BibitemOpen
  \bibfield  {author} {\bibinfo {author} {\bibfnamefont {N.}~\bibnamefont
  {Ruiz-Pino}}\ and\ \bibinfo {author} {\bibfnamefont {A.}~\bibnamefont
  {Prados}},\ }\bibfield  {title} {\bibinfo {title} {Markovian description of a
  wide class of feedback-controlled systems: application to the feedback
  flashing ratchet},\ }\href {https://doi.org/10.1088/1742-5468/ad64bb}
  {\bibfield  {journal} {\bibinfo  {journal} {Journal of Statistical Mechanics:
  Theory and Experiment}\ }\textbf {\bibinfo {volume} {2024}},\ \bibinfo
  {pages} {083204} (\bibinfo {year} {2024})}\BibitemShut {NoStop}%
\bibitem [{Note3()}]{Note3}%
  \BibitemOpen
  \bibinfo {note} {We take $k_B=1$ throughout the paper.}\BibitemShut {Stop}%
\bibitem [{Note4()}]{Note4}%
  \BibitemOpen
  \bibinfo {note} {\textcolor {black}{The variable $c$ may be multidimensional,
  we only assume that its set of possible values is countable.}}\BibitemShut
  {Stop}%
\bibitem [{SM-()}]{SM-Markovian-entropy-balance-2025}%
  \BibitemOpen
  \href@noop {} {}\bibinfo {note} {See Supplemental Material at [URL will be
  inserted by publisher] for more details of the Markovian description,
  detailed derivations of Eqs.~\eqref{eq:deltaS-FP-and-m},
  \eqref{eq:W-chain-bound}, \eqref{eq:comp-ineq}, \eqref{eq:W-Ix-Iu-equality},
  and \eqref{eq:W-Ix-Iu-equality-with-f}, and further numerical details and
  results.}\BibitemShut {Stop}%
\bibitem [{\citenamefont {Risken}(1996)}]{risken_fokker-planck_1996}%
  \BibitemOpen
  \bibfield  {author} {\bibinfo {author} {\bibfnamefont {H.}~\bibnamefont
  {Risken}},\ }\href {https://books.google.es/books?id=MG2V9vTgSgEC} {\emph
  {\bibinfo {title} {The {Fokker}-{Planck} {Equation}: {Methods} of {Solution}
  and {Applications}}}},\ Springer {Series} in {Synergetics}\ (\bibinfo
  {publisher} {Springer},\ \bibinfo {address} {Berlin Heidelberg},\ \bibinfo
  {year} {1996})\BibitemShut {NoStop}%
\bibitem [{\citenamefont {Gardiner}(2009)}]{gardiner_stochastic_2009}%
  \BibitemOpen
  \bibfield  {author} {\bibinfo {author} {\bibfnamefont {C.}~\bibnamefont
  {Gardiner}},\ }\href@noop {} {\emph {\bibinfo {title} {Stochastic {Methods}:
  {A} {Handbook} for the {Natural} and {Social} {Sciences}}}},\ Springer
  {Series} in {Synergetics}\ (\bibinfo  {publisher} {Springer Berlin
  Heidelberg},\ \bibinfo {address} {Berlin - Heidelberg},\ \bibinfo {year}
  {2009})\BibitemShut {NoStop}%
\bibitem [{Note5()}]{Note5}%
  \BibitemOpen
  \bibinfo {note} {We follow the standard convention for heat and work, both of
  them are positive (negative) when they lead to an increase (decrease) of the
  internal energy of the system.}\BibitemShut {Stop}%
\bibitem [{\citenamefont {Cover}\ and\ \citenamefont
  {Thomas}(2006)}]{cover_elements_2006}%
  \BibitemOpen
  \bibfield  {author} {\bibinfo {author} {\bibfnamefont {T.~M.}\ \bibnamefont
  {Cover}}\ and\ \bibinfo {author} {\bibfnamefont {J.~A.}\ \bibnamefont
  {Thomas}},\ }\href@noop {} {\emph {\bibinfo {title} {Elements of information
  theory}}},\ \bibinfo {edition} {2nd}\ ed.\ (\bibinfo  {publisher}
  {Wiley-Interscience},\ \bibinfo {address} {Hoboken, N.J},\ \bibinfo {year}
  {2006})\BibitemShut {NoStop}%
\bibitem [{Note6()}]{Note6}%
  \BibitemOpen
  \bibinfo {note} {\textcolor {black}{For spatially periodic potentials,} the
  force $f$ breaks detailed balance. $J_{\protect \text {s}}(x|c)\textcolor
  {black}{\protect \ne 0}$ is independent of $x$ \textcolor {black}{for $d=1$},
  \textcolor {black}{whereas it} may depend on $x$ \textcolor {black}{for
  $d>1$}.}\BibitemShut {Stop}%
\bibitem [{\citenamefont {Oono}\ and\ \citenamefont
  {Paniconi}(1998)}]{oono_steady_1998}%
  \BibitemOpen
  \bibfield  {author} {\bibinfo {author} {\bibfnamefont {Y.}~\bibnamefont
  {Oono}}\ and\ \bibinfo {author} {\bibfnamefont {M.}~\bibnamefont
  {Paniconi}},\ }\bibfield  {title} {\bibinfo {title} {Steady {State}
  {Thermodynamics}},\ }\href@noop {} {\bibfield  {journal} {\bibinfo  {journal}
  {Progress of Theoretical Physics Supplement}\ }\textbf {\bibinfo {volume}
  {130}},\ \bibinfo {pages} {29} (\bibinfo {year} {1998})}\BibitemShut
  {NoStop}%
\bibitem [{\citenamefont {Hatano}\ and\ \citenamefont
  {Sasa}(2001)}]{hatano_steady-state_2001}%
  \BibitemOpen
  \bibfield  {author} {\bibinfo {author} {\bibfnamefont {T.}~\bibnamefont
  {Hatano}}\ and\ \bibinfo {author} {\bibfnamefont {S.-i.}\ \bibnamefont
  {Sasa}},\ }\bibfield  {title} {\bibinfo {title} {Steady-{State}
  {Thermodynamics} of {Langevin} {Systems}},\ }\href
  {https://doi.org/10.1103/PhysRevLett.86.3463} {\bibfield  {journal} {\bibinfo
   {journal} {Physical Review Letters}\ }\textbf {\bibinfo {volume} {86}},\
  \bibinfo {pages} {3463} (\bibinfo {year} {2001})}\BibitemShut {NoStop}%
\bibitem [{\citenamefont {Speck}\ and\ \citenamefont
  {Seifert}(2005)}]{speck_integral_2005}%
  \BibitemOpen
  \bibfield  {author} {\bibinfo {author} {\bibfnamefont {T.}~\bibnamefont
  {Speck}}\ and\ \bibinfo {author} {\bibfnamefont {U.}~\bibnamefont
  {Seifert}},\ }\bibfield  {title} {\bibinfo {title} {Integral fluctuation
  theorem for the housekeeping heat},\ }\href
  {https://doi.org/10.1088/0305-4470/38/34/L03} {\bibfield  {journal} {\bibinfo
   {journal} {Journal of Physics A: Mathematical and General}\ }\textbf
  {\bibinfo {volume} {38}},\ \bibinfo {pages} {L581} (\bibinfo {year}
  {2005})}\BibitemShut {NoStop}%
\bibitem [{\citenamefont {Van~den Broeck}\ and\ \citenamefont
  {Esposito}(2010)}]{van_den_broeck_three_2010}%
  \BibitemOpen
  \bibfield  {author} {\bibinfo {author} {\bibfnamefont {C.}~\bibnamefont
  {Van~den Broeck}}\ and\ \bibinfo {author} {\bibfnamefont {M.}~\bibnamefont
  {Esposito}},\ }\bibfield  {title} {\bibinfo {title} {Three faces of the
  second law. {II}. {Fokker}-{Planck} formulation},\ }\href
  {https://doi.org/10.1103/PhysRevE.82.011144} {\bibfield  {journal} {\bibinfo
  {journal} {Physical Review E}\ }\textbf {\bibinfo {volume} {82}},\ \bibinfo
  {pages} {011144} (\bibinfo {year} {2010})}\BibitemShut {NoStop}%
\bibitem [{\citenamefont {Esposito}\ and\ \citenamefont {Van
  Den~Broeck}(2010)}]{esposito_three_2010}%
  \BibitemOpen
  \bibfield  {author} {\bibinfo {author} {\bibfnamefont {M.}~\bibnamefont
  {Esposito}}\ and\ \bibinfo {author} {\bibfnamefont {C.}~\bibnamefont {Van
  Den~Broeck}},\ }\bibfield  {title} {\bibinfo {title} {Three {Detailed}
  {Fluctuation} {Theorems}},\ }\href
  {https://doi.org/10.1103/PhysRevLett.104.090601} {\bibfield  {journal}
  {\bibinfo  {journal} {Physical Review Letters}\ }\textbf {\bibinfo {volume}
  {104}},\ \bibinfo {pages} {090601} (\bibinfo {year} {2010})}\BibitemShut
  {NoStop}%
\bibitem [{Note7()}]{Note7}%
  \BibitemOpen
  \bibinfo {note} {\textcolor {black}{For error-free feedback control,
  Eq.~\protect \eqref {eq:Ic-def} reduces to $\protect \overline {\protect
  \mathcal {I}_{\protect \vec {c}}}=S_{\protect \vec {c}}(t_n^+)-S_{\protect
  \vec {c}}(t_n^-)=S_{c_n|\protect \vec {c}_{n-1}}\ge 0$, see also~\cite
  {ruiz-pino_information_2023}.}}\BibitemShut {Stop}%
\bibitem [{Note8()}]{Note8}%
  \BibitemOpen
  \bibinfo {note} {\textcolor {black}{This choice preserves the physical image
  of a sinusoidal potential~\cite {toyabe_experimental_2010} while simplifying
  analytical calculations.}}\BibitemShut {Stop}%
\bibitem [{\citenamefont {Archambault}\ \emph {et~al.}(2024)\citenamefont
  {Archambault}, \citenamefont {Crauste-Thibierge}, \citenamefont {Ciliberto},\
  and\ \citenamefont {Bellon}}]{archambault_inertial_2024}%
  \BibitemOpen
  \bibfield  {author} {\bibinfo {author} {\bibfnamefont {A.}~\bibnamefont
  {Archambault}}, \bibinfo {author} {\bibfnamefont {C.}~\bibnamefont
  {Crauste-Thibierge}}, \bibinfo {author} {\bibfnamefont {S.}~\bibnamefont
  {Ciliberto}},\ and\ \bibinfo {author} {\bibfnamefont {L.}~\bibnamefont
  {Bellon}},\ }\bibfield  {title} {\bibinfo {title} {Inertial effects in
  discrete sampling information engines},\ }\href
  {https://doi.org/10.1209/0295-5075/ad8bf0} {\bibfield  {journal} {\bibinfo
  {journal} {Europhysics Letters}\ }\textbf {\bibinfo {volume} {148}},\
  \bibinfo {pages} {41002} (\bibinfo {year} {2024})}\BibitemShut {NoStop}%
\bibitem [{\citenamefont {Dago}\ \emph {et~al.}(2021)\citenamefont {Dago},
  \citenamefont {Pereda}, \citenamefont {Barros}, \citenamefont {Ciliberto},\
  and\ \citenamefont {Bellon}}]{dago_information_2021}%
  \BibitemOpen
  \bibfield  {author} {\bibinfo {author} {\bibfnamefont {S.}~\bibnamefont
  {Dago}}, \bibinfo {author} {\bibfnamefont {J.}~\bibnamefont {Pereda}},
  \bibinfo {author} {\bibfnamefont {N.}~\bibnamefont {Barros}}, \bibinfo
  {author} {\bibfnamefont {S.}~\bibnamefont {Ciliberto}},\ and\ \bibinfo
  {author} {\bibfnamefont {L.}~\bibnamefont {Bellon}},\ }\bibfield  {title}
  {\bibinfo {title} {Information and {Thermodynamics}: {Fast} and {Precise}
  {Approach} to {Landauer}’s {Bound} in an {Underdamped} {Micromechanical}
  {Oscillator}},\ }\href {https://doi.org/10.1103/PhysRevLett.126.170601}
  {\bibfield  {journal} {\bibinfo  {journal} {Physical Review Letters}\
  }\textbf {\bibinfo {volume} {126}},\ \bibinfo {pages} {170601} (\bibinfo
  {year} {2021})}\BibitemShut {NoStop}%
\bibitem [{\citenamefont {Dago}\ and\ \citenamefont
  {Bellon}(2022)}]{dago_dynamics_2022}%
  \BibitemOpen
  \bibfield  {author} {\bibinfo {author} {\bibfnamefont {S.}~\bibnamefont
  {Dago}}\ and\ \bibinfo {author} {\bibfnamefont {L.}~\bibnamefont {Bellon}},\
  }\bibfield  {title} {\bibinfo {title} {Dynamics of {Information} {Erasure}
  and {Extension} of {Landauer}’s {Bound} to {Fast} {Processes}},\ }\href
  {https://doi.org/10.1103/PhysRevLett.128.070604} {\bibfield  {journal}
  {\bibinfo  {journal} {Physical Review Letters}\ }\textbf {\bibinfo {volume}
  {128}},\ \bibinfo {pages} {070604} (\bibinfo {year} {2022})}\BibitemShut
  {NoStop}%
\bibitem [{\citenamefont {Dago}\ \emph {et~al.}(2023)\citenamefont {Dago},
  \citenamefont {Ciliberto},\ and\ \citenamefont
  {Bellon}}]{dago_adiabatic_2023}%
  \BibitemOpen
  \bibfield  {author} {\bibinfo {author} {\bibfnamefont {S.}~\bibnamefont
  {Dago}}, \bibinfo {author} {\bibfnamefont {S.}~\bibnamefont {Ciliberto}},\
  and\ \bibinfo {author} {\bibfnamefont {L.}~\bibnamefont {Bellon}},\
  }\bibfield  {title} {\bibinfo {title} {Adiabatic computing for optimal
  thermodynamic efficiency of information processing},\ }\href
  {https://doi.org/10.1073/pnas.2301742120} {\bibfield  {journal} {\bibinfo
  {journal} {Proceedings of the National Academy of Sciences}\ }\textbf
  {\bibinfo {volume} {120}},\ \bibinfo {pages} {e2301742120} (\bibinfo {year}
  {2023})}\BibitemShut {NoStop}%
\bibitem [{git()}]{github-page-RyP25}%
  \BibitemOpen
  \href@noop {} {}\bibinfo {howpublished}
  {\url{https://github.com/fine-group-us/Markovian_entropy_balance}}\BibitemShut
  {NoStop}%
\bibitem [{Note9()}]{Note9}%
  \BibitemOpen
  \bibinfo {note} {For $\Delta t_m\to \infty $ and $\Delta x=0$, the values for
  $\beta \left \langle W \right \rangle =\protect \overline {\protect \mathcal
  {I}_u - \protect \mathcal {I}_{\protect \vec {x}}}$ coincide with those
  reported in Ref.~\cite {paneru_lossless_2018}.}\BibitemShut {Stop}%
\end{thebibliography}%

{\begin{center}
  {\bf End Matter}  
\end{center}}

\textit{Lossless harmonic engine--} Here, we illustrate our results in a different information engine, inspired by a ``lossless'' information engine~\cite{paneru_lossless_2018,archambault_inertial_2024}. Our setup is very close to the experimental implementation in Ref.~\cite{archambault_inertial_2024}. We consider a Brownian particle confined by a harmonic trap of stiffness $k$, the centre of which fluctuates between $-L$ and $+L$, i.e.~the particle feels a fluctuating potential $V(x,c)=k(x-cL)^2/2$, with $c=\pm 1$. Units are the same as in the example in the main text.

\begin{figure}[!hb]
    \centering
\includegraphics[width=3.25in]{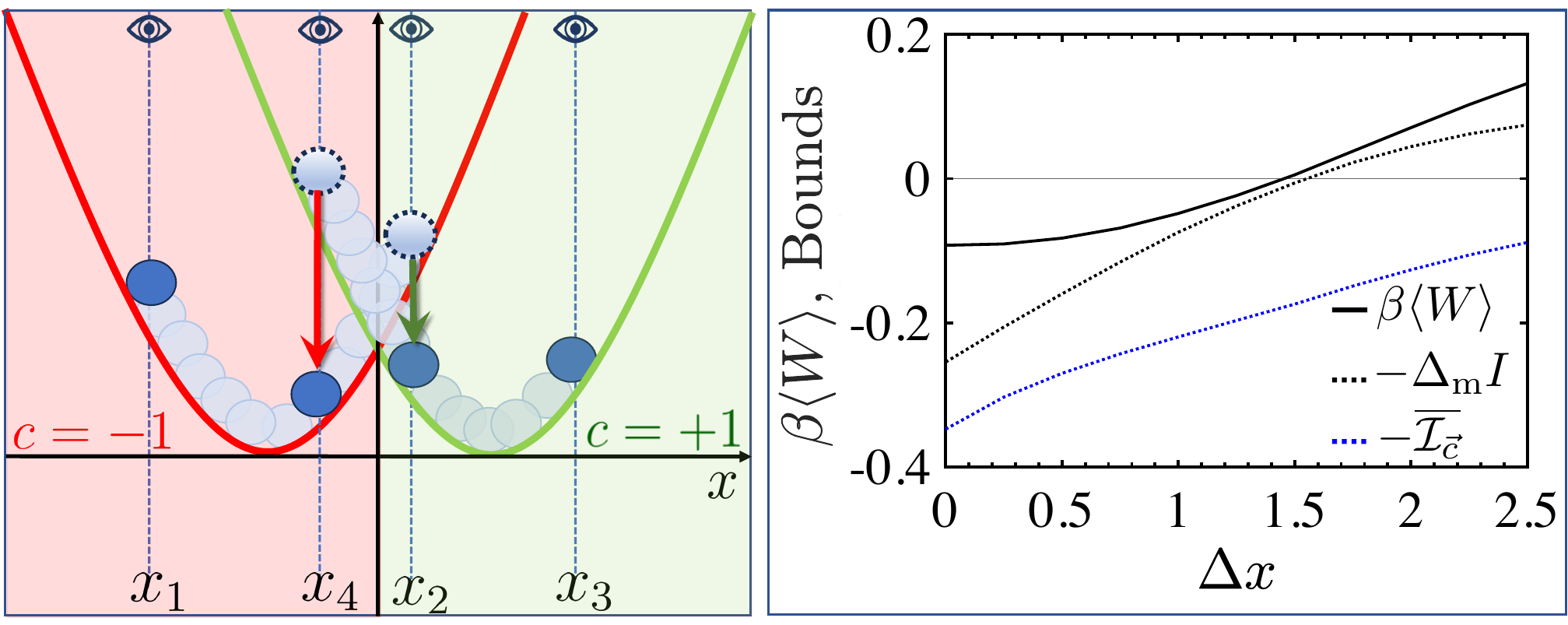}
    \caption{(Left) Sketch of the harmonic information engine. The centre of the trap may be switched between $-L$ and $+L$ by the feedback controller after measuring the position of the particle $x_n$ at times $t_n=n\Delta t_{\meas}$. (Right) Average work and bounds as a function of the uncertainty in the measurement $\Delta x$. Note that $\Delta x$ has no upper bound here because $x\in(-\infty,+\infty)$. Additional parameters are $\Delta t_{\meas}=0.5$ and  $k=1$. }
    \label{fig:engine-harmonic-bounds}
\end{figure}

For perfect measurement, the control variable is set to $c_n=-1$ ($c_n=+1$) and thus activates the trap centred at $-L$ ($+L$) if the measured position at time $t_n$ verifies $x_n<0$ ($x_n>0$)---as shown in the left panel of Fig.~\ref{fig:engine-harmonic-bounds}. Since the two potentials intersect at $x=0$, a perfect protocol
always extracts work at the measurement. Again, we consider a uniform measurement error, $\Delta x$: the position of the particle is known to be uniformly distributed in the interval $(x-\Delta x,x+\Delta x)$. Therefore, the controller sets $c_n=-1$ ($c_n=+1$) with probability proportional to the length of the subinterval of $(x-\Delta x,x+\Delta x)$ that is to the left (right) of the origin. 

Here, natural boundary conditions apply, $x \in \calD=\mathbb{R}$, in contrast to the periodic boundaries of the example in the main text. Thus, an external force $f$ does not break detailed balance, the steady distribution $\Psxfc$ is the canonical one for $\tilde{V}(x,c)=V(x,c)-fx$. For harmonic confinement, this simply corresponds to a shift of the trap's centre---we thus restrict ourselves to $f=0$. 

In the right panel of Fig.~\eqref{fig:engine-harmonic-bounds}, we observe that, as predicted by Eq.~\eqref{eq:comp-ineq-simple}, the Markovian bound beats the chain one for all values of $\Delta x$. The improvement provided by the Markovian bound is significant throughout and, as in the model engine in the main text, the novel bound accurately predicts the value of the error above which the information engine ceases to be useful because $\av{W}>0$.

\begin{figure}
    \centering
\includegraphics[width=3.25in]{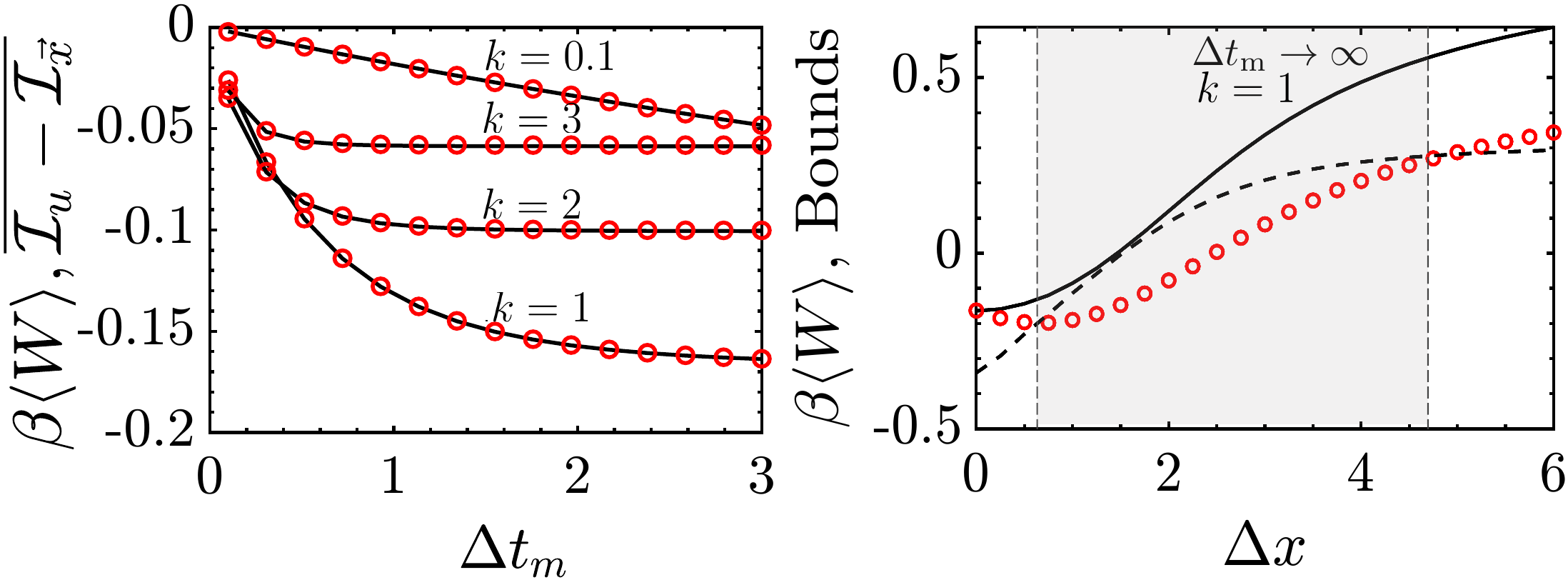}
    \caption{Same as in Fig.~\ref{fig:W-I-Iu}, for the harmonic information engine. In the left panel, the transition from the low-temperature to the high-temperature regime is now given by the value of $k$, from $k=0.1$ ($k\ll T$, high temperature) to $k=3$ ($k\gg  T$, low temperature). The behaviour found is analogous to that in Fig.~\ref{fig:W-I-Iu}, showing the robustness of the observed qualitative picture for different information engines.
    \label{fig:W-I-Iu-harmonic}
}
\end{figure}

The tightness of the unavailable information bound for error-free measurement, $\Delta x=0$,  is checked in the left panel of Fig.~\ref{fig:W-I-Iu-harmonic}. Specifically, we do so for several values of the stiffness, corresponding to low-, intermediate-, and high-temperature regimes. Again, $\av{W}$ exhibits a non-monotonic dependence on $k$, since it vanishes both for $k\to 0^+$ and for $k\to +\infty$. This is physically reasonable: for $k\to 0^+$, we have no trap and thus no potential energy, whereas for $k\to+\infty$ the two traps become infinitely confining and switching with perfect measurement is completely prevented. Note that the unavailable information bound is tight for all values of $\Delta t_m$, as predicted by our Eq.~\eqref{eq:W-Ix-Iu-equality}: it is not necessary to let the particle completely relax between measurements, as done in Refs.~\cite{paneru_lossless_2018,archambault_inertial_2024}, to get a ``lossless'' information engine.

For imperfect measurements in the limit $\Delta t_{\meas}\gg 1$, we compare the unavailable information bound---which is no longer tight---and the Markovian bound in the right panel of Fig.~\ref{fig:W-I-Iu-harmonic}. Again, the Markovian bound outperforms the unavailable information bound in a certain region, which includes the value $\Delta x_0$ at which $\av{W}$ changes sign~\footnote{For $\Delta t_m\to\infty$ and $\Delta x=0$, the values for $\beta\av{W}=\overline{\calI_u - \calI_{\vec{x}}}$ coincide with those reported in Ref.~\cite{paneru_lossless_2018}.}.

\end{document}